\documentclass[12pt]{article}
\usepackage{amsmath}
\usepackage{graphicx}
\usepackage{enumerate}

\newcommand{\blind}{1}

\usepackage[utf8]{inputenc} 
\usepackage[T1]{fontenc}    
\usepackage{hyperref}
\hypersetup{
  urlcolor     = red, 
  linkcolor    = brown, 
  citecolor   = teal 
}
\usepackage{url}            
\usepackage{booktabs}       
\usepackage{amsfonts}       
\usepackage{nicefrac}       
\usepackage{microtype}      
\usepackage{lipsum}
\usepackage{amsmath}
\usepackage{xcolor}
\usepackage{amssymb}
\usepackage{multicol}
\usepackage{pdflscape}
\usepackage{setspace}
\usepackage{fancyvrb}

\usepackage{centernot}
\usepackage{comment}
\usepackage[round]{natbib}
\usepackage{enumitem}
\setenumerate{noitemsep,topsep=0pt,parsep=0pt,partopsep=0pt}
\setitemize{noitemsep,topsep=0pt,parsep=0pt,partopsep=0pt}
\usepackage{afterpage}
\usepackage{tablefootnote}
\usepackage{tabularx,booktabs} 
\usepackage{lscape}
\usepackage{multirow}

\usepackage{fancyhdr}

\usepackage{amsthm}
\theoremstyle{definition}

\usepackage{thmtools}
\usepackage{thm-restate}
\usepackage{cleveref}

\usepackage{color}
\usepackage{tikz}
\usetikzlibrary{shapes,decorations,arrows,calc,arrows.meta,fit,positioning}
\tikzset{
    -Latex,auto,node distance =1 cm and 1 cm,semithick,
    state/.style ={ellipse, draw, minimum width = 0.7 cm},
    point/.style = {circle, draw, inner sep=0.04cm,fill,node contents={}},
    bidirected/.style={Latex-Latex,dashed},
    el/.style = {inner sep=2pt, align=left, sloped}
}
\usepackage{helvet}
\usepackage{hyperref}
\usepackage[utf8]{inputenc}
\usepackage[english]{babel}

\DeclareMathOperator{\sd}{sd}
\DeclareMathOperator{\se}{se}

\begin{document}

\def\spacingset#1{\renewcommand{\baselinestretch}%
{#1}\small\normalsize} \spacingset{1}

\if1\blind
{
  \title{\bf Causal progress with imperfect placebo treatments and outcomes}
  \author{Adam Rohde\thanks{Decision Science Technical Lead, ZipRecruiter. adamrohde@g.ucla.edu}, Chad Hazlett\footnote{Professor, Department of Statistics \& Data Science; Department of Political Science, UCLA. chazlett@ucla.edu}}
  \maketitle
} \fi

\if0\blind
{
  \bigskip
  \bigskip
  \bigskip
  \begin{center}
    {\LARGE\bf Causal progress with imperfect placebo treatments and outcomes}
\end{center}
  \medskip
} \fi

\bigskip
\begin{abstract}
In the quest to make defensible causal claims from observational data, researchers may leverage information from “placebo treatments” and “placebo outcomes”. Existing approaches assume (i) perfect placebos (meaning placebo treatments have no effect on outcomes and the real treatment has no effect on placebo outcomes) and (ii) equiconfounding (equal confounding across placebo and real treatment-outcome relationships). Many also require the availability of a valid placebo treatment and placebo outcome. We propose a new framework based on omitted variable bias, allowing users to postulate ranges of values corresponding to unequal confounding and imperfect placebos. These assumptions identify or bound linear treatment effect estimates, with either a placebo outcome or treatment. While applicable in many settings, one case for this method is the use of pre-treatment outcomes as placebo outcomes, as in difference-in-difference. In this setting, the parallel trends assumption is identical to equiconfounding on a specific scale, to which our framework provides a reasoned relaxation. We demonstrate this approach in applications that examine the effect of a job training program, and another that estimates the effect of the Zika virus on birth rates in Brazil. These analyses employ the \texttt{R} package \texttt{placeboLM} package.
\end{abstract}

\noindent%
\vfill

\newpage

\onehalfspacing

\section{Introduction}

Across disciplines, unobserved confounding plagues researchers seeking to draw causal inferences from observational data. When we cannot rule out unobserved confounding, we may look to alternative strategies to fully or partially identify the range of defensible causal effects given the assumptions we can support. One set of approaches rests on observing variables that (i) may be affected by similar factors as those affecting the treatment, but which cannot affect the outcome (placebo treatments), and/or (ii) variables sharing similar influences with the outcome, but that cannot be affected by treatment (placebo outcomes).\footnote{In related literatures such as the proximal causal inference literature discussed more below, the terms ``negative control outcome'' and ``negative control exposure'' have been used to reference ``placebo outcome'' and ``placebo treatment'', respectively.} The broad intuition is then that, if these variables are indeed similar to their real counterparts in terms of the degree of confounding they suffer, they allow us to estimate and remove that much confounding from the real treatment-outcome relationship. 

With recent exceptions we describe below, the primary framework for exploiting this idea has focused on the goal of point identification, and relies on two point assumptions. First, the placebo treatments/outcomes involved must be ``perfect placebos'', meaning there must be precisely zero effect of the placebo treatment on the real outcome, or real treatment on placebo outcome. Second, these approaches must connect the amount of confounding suffered by the placebo relationship to that in the relationship of interest, typically by assuming they are equal on some scale. We call this the  ``equiconfounding'' assumption. As discussed below, such approaches include those described in recent work on proximal causal inference \citep{Miao2020, Tchetgen2020}, which in its original form, requires a perfect placebo treatment, perfect placebo outcome, and a ``bridge function'' assumption that invokes equiconfounding on some transformation of the variables. We discuss this and a related approaches below. 

Unfortunately, when seeking to defend causal claims in real world circumstances, these assumptions are often not defensible. While some placebos may be arguably ``perfect'' (such as pre-treatment outcome measures acting as placebo outcomes), for many other variables one might hope to regard as a placebo outcome or treatment, this is not the case. More problematic still is the  equiconfounding assumption. On what basis can investigators hope to argue that unobserved confounders have precisely the same impact on the treatment-outcome relationship as in the relationship with the placebo treatment or outcome? 

Is it nevertheless possible to make use of the information that a given variable is a possibly imperfect placebo, and is exposed to some fraction of the confounding that affect the true treatment-outcome relationship? In this paper we examine what progress can be made by sidestepping the two strong assumptions of placebo perfection and equiconfounding, showing how arguments about limited violations of these assumptions lead to defensible ranges of causal claims. Specifically, consider a set of interval assumptions about (i) the relative confounding felt by the treatment-outcome relationship compared to that in the placebo treatment to real outcome or real treatment to placebo outcome relationship; and (ii) the degree of ``imperfection'' in each placebo.  For any range of assumptions on these that can(not) be ruled out by argumentation, there is a set of effect estimates that consequently can(not) be ruled out as consistent with the evidence. This approach simply shows the consequence of any assumptions one would be willing to pose regarding these placebo relationships. One can expect relatively uninformative results when little is known about the extent to which these assumptions are violated, whereas the approach will be more informative to the degree that investigators are better able to defend narrower assumptions about the degree of placebo imperfection or the relative strength of confounding.

In what follows, we begin by discussing causal structures in general that involve placebo treatments and outcomes, including cases where these labels become ambiguous. Our treatment employs linear models in an omitted variable bias framework. We focus first on a single placebo outcome, then show the corresponding analysis for placebo treatments.\footnote{We reserve our analysis of ``double placebos''---cases with a placebo treatment and outcome---for the appendix, noting that this relates most closely to developments under the heading of proximal causal inference.} 
We give special attention to the connection of this approach to difference-in-difference (DID), in which (i) the pre-treatment outcome serves as a (perfect) placebo outcome, and (ii) the parallel trends assumption is precisely the equiconfounding assumption, on the additive scale. This relationship is clarifying, but also demonstrates the our approach offers one relaxation of DID, enabling investigators to examine the credible causal interval obtained over a set of assumptions (about relative confounding) rather than the point estimate given under the point assumption of parallel trends. 

After a simulation example, we apply this approach to two real-world investigations. First, we use data from the well-known National Supported Work Demonstration to show that under reasonable  assumptions an investigator given the observational data would have been able to produce a range of estimates that contains the experimental benchmark value. 
Second, we show that despite apparent evidence to the contrary, the Zika virus likely decreased birth rates in Brazil in 2016. We conduct these analyses using the \texttt{PlaceboLM} package, which we make available in the \texttt{R} computing language.

\section{Leveraging placebos in linear models}
\label{proposal}

\subsection{Causal structures of interest}

Placebo treatments and outcomes, perfect or imperfect, are defined by the role they play in some causal model. We employ directed acyclic graphs (DAGs) \citep{Pearl1995} to label these roles. Figure \ref{f_simple_dags} shows simple structures with ``perfect'' placebo treatments ($P$) and placebo outcomes ($N$).  Placebo treatments and outcomes can look very similar in their causal role; indeed Figure \ref{f_simple_dags}(a) and (b) are mirror images, replacing the role of $P$ with $N$. 
We explore such ambiguities in the context of a broader set of cases in the Online Supplement.
To call a placebo ``imperfect'' is to allow for an additional causal effect, and so always implies an additional edge in the causal structure. For example, Figure~\ref{f_simple_dags}(c) adds the $D \to N$ edge to Figure \ref{f_simple_dags}(a), while Figure~\ref{f_simple_dags}(d) adds the $P\to Y$ edge to Figure \ref{f_simple_dags}(b).

\begin{figure}[h!]
\begin{center}
\begin{tikzpicture}[baseline=(current bounding box.north)]
\node (0) at (1.5,3) {(a)};
\node (D) at (0,0) {$D$};
\node (Y) at (2,0) {$Y$};
\node (N) at (2,1) {$N$};
\node (Z) at (0,2) {$\mathbf{Z}$};
\node[gray] (X) at (3,1) {$\mathbf{X}$};
\path (D) edge (Y);
\path (Z) edge (D);
\path (Z) edge (Y);
\path (Z) edge (N);
\path[gray] (X) edge [out=270,in=310] (D);
\path[gray] (X) edge (Y);
\path[gray] (X) edge (N);
\path[gray,bidirected] (X) edge[bend right=40] (Z);
\end{tikzpicture}\hspace{1cm}
\begin{tikzpicture}[baseline=(current bounding box.north)]
\node (0) at (1.5,3) {(b)};
\node (D) at (0,0) {$D$};
\node (Y) at (2,0) {$Y$};
\node (P) at (0,1) {$P$};
\node (Z) at (2,2) {$\mathbf{Z}$};
\node[gray] (X) at (-1,1) {$\mathbf{X}$};
\path (D) edge (Y);
\path (Z) edge (D);
\path (Z) edge (Y);
\path (Z) edge (P);
\path[gray] (X) edge [out=270,in=230] (Y);
\path[gray] (X) edge (D);
\path[gray] (X) edge (P);
\path[gray,bidirected] (X) edge[bend left=40] (Z);
\end{tikzpicture}\hspace{1cm}
\begin{tikzpicture}[baseline=(current bounding box.north)]
\node (0) at (1.5,3) {(c)};
\node (D) at (0,0) {$D$};
\node (Y) at (2,0) {$Y$};
\node (N) at (2,1) {$N$};
\node (Z) at (0,2) {$\mathbf{Z}$};
\node[gray] (X) at (3,1) {$\mathbf{X}$};
\path (D) edge (Y);
\path (Z) edge (D);
\path (Z) edge (Y);
\path (Z) edge (N);
\path[gray] (X) edge [out=270,in=310] (D);
\path[gray] (X) edge (Y);
\path[gray] (X) edge (N);
\path[gray,bidirected] (X) edge[bend right=40] (Z);
\path (D) edge (N);
\end{tikzpicture}\hspace{1cm}
\begin{tikzpicture}[baseline=(current bounding box.north)]
\node (0) at (1.5,3) {(d)};
\node (D) at (0,0) {$D$};
\node (Y) at (2,0) {$Y$};
\node (P) at (0,1) {$P$};
\node (Z) at (2,2) {$\mathbf{Z}$};
\node[gray] (X) at (-1,1) {$\mathbf{X}$};
\path (D) edge (Y);
\path (Z) edge (D);
\path (Z) edge (Y);
\path (Z) edge (P);
\path[gray] (X) edge [out=270,in=230] (Y);
\path[gray] (X) edge (D);
\path[gray] (X) edge (P);
\path[gray,bidirected] (X) edge[bend left=40] (Z);
\path (P) edge (Y);
\end{tikzpicture}
\caption{\label{f_simple_dags} Simple DAGs with a placebo outcome, $N$, and with a placebo treatment, $P$. $D$ is treatment, $Y$ is outcome, $\mathbf{Z}$ contains unobserved confounders, and $\mathbf{X}$ contains observed covariates.}
\end{center}
\end{figure}

\subsection{An omitted variable bias approach to placebo outcomes}
\label{proposal_po}

Our proposed framework is a product of the omitted variable bias analysis of linear models (e.g. \citealp{angrist_mostly_2008,CinelliHazlett2020}), extended to make use of assumptions involving placebo treatments or outcomes.\footnote{We also consider relaxations to this parametric approach in Appendix \ref{PLandNPAppendix}, in which we illustrate similar methods adapted to partially linear and non-parametric estimation of treatment effects.} 
We begin by discussing placebo outcomes, and build on this in the subsequent analysis of placebo treatments and of ``double placebos'' in the appendix. 
Suppose that we want to estimate the ``long'' regression of $Y$ on $D$, $\mathbf{X}$, and $\mathbf{Z}$ in Equation \ref{e_long_reg_outcome_y}. In particular, we are interested in the coefficient $\beta_{Y\sim D | Z,X}$ as a linear approximation to the effect of $D$ on $Y$. $\mathbf{X}$ and $\mathbf{Z}$ are vectors of covariates.\footnote{Note that in this section and others, we will refer to including $\mathbf{Z}$ in the long regression as ``de-confounding''. Readers should recognize that, in reality, we are only partialling out the linear relationships with $\mathbf{Z}$. This is, therefore, a linear approximation to full non-parametric elimination of confounding from $\mathbf{Z}$. On the other hand, $\mathbf{Z}$ can be assumed to contain arbitrary functions of the relevant variables, making this very flexible for de-confounding.}
\begin{equation} \label{e_long_reg_outcome_y}
\begin{aligned}
Y &= \beta_{Y\sim D| Z, X} D + \mathbf{X}\beta_{Y \sim X| D,Z} + \mathbf{Z}\beta_{Y \sim Z|D,X}   + \epsilon_l
\end{aligned}
\end{equation}
In reality, however, the variables $\mathbf{Z}$ are not observed, and henceforth are referred to as ``unobserved confounders''. We are restricted to estimating the ``short'' regression of $Y$ on $D$ and $\mathbf{X}$ in Equation \ref{e_short_reg_outcome_y}.
\begin{equation} \label{e_short_reg_outcome_y}
\begin{aligned}
Y &= \beta_{Y\sim D | X} D + \mathbf{X}\beta_{Y \sim X| D} + \epsilon_s
\end{aligned}
\end{equation}

The difference between $\beta_{Y\sim D | X}$ and $\beta_{Y\sim D| Z, X}$ is the bias due to the omission of $\mathbf{Z}$. As in \cite{CinelliHazlett2020}, we are interested in sample regressions that are actually run and how the coefficient estimate obtained differs from the coefficient estimate that would have been obtained in the same sample had it been possible to include $\mathbf{Z}$. 
We label the bias due to the omission of $\mathbf{Z}$ in the sample regression as 
\begin{equation} \label{e_bias_y_trad_ovb}
\begin{aligned}
\widehat{\text{bias}}_{\text{(YD.X)}}
\overset{\Delta}{=} 
\widehat{\beta}_{Y\sim D | X} - \widehat{\beta}_{Y\sim D| Z, X}  
\end{aligned}
\end{equation}

Our aim is to show how the observed relationship between the treatment and placebo outcome, together with postulated assumptions, lead to estimates of this bias that can then be removed from the initial estimate. For example, if an investigator argues that the relationship between the treatment and the placebo outcome (the observed $D\to N$ relationship) suffers from a ``similar'' level of unobserved confounding as does the relationship between treatment and real outcome (the observed $D\to Y$ relationship), how could we make use of this information? 

To do so we begin by considering the ``long'' regression of $N$ on $D$, $\mathbf{X}$ and $\mathbf{Z}$ (Equation~\ref{e_long_reg_outcome_n}),
\begin{equation} \label{e_long_reg_outcome_n}
\begin{aligned}
N &= \beta_{N\sim D| Z, X} D + \mathbf{X}\beta_{N \sim X| D,Z} + \mathbf{Z}\beta_{N \sim Z|D,X} + \xi_l
\end{aligned}
\end{equation}

\noindent which we cannot estimate. Note that $\mathbf{Z}$ here represents a rich enough set of variables that it de-confounds both the $D\to Y$ and the $D\to N$ relationships. We can instead however estimate the short regression of $N$ on just $\mathbf{X}$ and $D$ (Equation~\ref{e_short_reg_outcome_n}), 
\begin{equation} \label{e_short_reg_outcome_n}
\begin{aligned}
N &= \beta_{N\sim D | X} D + \mathbf{X}\beta_{N \sim X| D} + \xi_s
\end{aligned}
\end{equation}

Here again we take interest in the difference between the coefficient estimate $\hat{\beta}_{N\sim D | X}$ that a researcher actually obtains in a sample regression, and the hypothetical coefficient $\hat{\beta}_{N\sim D| Z, X}$ that they would have obtained if $\mathbf{Z}$ were included. The differece is the sample omitted variable bias, 
\begin{equation} \label{e_bias_n_trad_ovb}
\begin{aligned}
\widehat{\text{bias}}_{\text{(ND.X)}}
\overset{\Delta}{=} 
 \hat{\beta}_{N\sim D | X} - \hat{\beta}_{N\sim D| Z, X}  
\end{aligned}
\end{equation}

The key maneuver for making this useful is to then rewrite the bias on the coefficient of interest due to the omission of $\mathbf{Z}$ as a function of the bias in the treatment-placebo relationship. A first way to do so is simply to define $m \in \mathbb{R}$ such that
\vspace{-.2in}
\begin{align}\label{mdef}
m \overset{\Delta}{=}  \frac{\widehat{\text{bias}}_{\text{(YD.X)}}}{\widehat{\text{bias}}_{\text{(ND.X)}}} 
\end{align}

\noindent The value of $m$ is not known in reality, but the supposition will be that we can postulate a range of values for it and see the consequence. For any postulated value of $m$, we can solve for $\hat{\beta}_{Y\sim D| Z, X}$ at that value of $m$ by algebraic rearrangement,
\vspace{-.2in}
\begin{align} 
\widehat{\text{bias}}_{\text{(YD.X)}}
&= \hat{\beta}_{Y\sim D | X} - \hat{\beta}_{Y\sim D| Z, X}  \\
&= m \times \left(\hat{\beta}_{N\sim D | X} - \hat{\beta}_{N\sim D| Z, X} \right) \\ 
\hat{\beta}_{Y\sim D| Z, X} &= \hat{\beta}_{Y\sim D | X} - m \times \left(\hat{\beta}_{N\sim D | X} - \hat{\beta}_{N\sim D| Z, X} \right) \label{e_revised_estimate_y_trad_ovb_reexpress}
\end{align}

Equation \ref{e_revised_estimate_y_trad_ovb_reexpress} provides all we need for partial identification of $\hat{\beta}_{Y\sim D| Z, X}$.  The data provide us with $\hat{\beta}_{Y \sim D|X}$ and $\hat{\beta}_{N \sim D |X}$, whereas $m$ and $\hat{\beta}_{N\sim D|Z,X}$ parameterize assumptions from outside the data, and can be postulated for purposes of partial identification or sensitivity analysis. 
Specifically, $\hat{\beta}_{N\sim D|Z,X}$ parameterizes a postulated violation of ``placebo perfection''; i.e. the $D \to N$ path in Figure \ref{f_simple_dags}(c).\footnote{We write $\hat{\beta}_{N\sim D|Z,X}$ with the ``hat'' because it is the value of this quantity in a given sample regression. However, it is an object of assumption (i.e. a sensitivity parameter), so arguments regarding it will arise from information outside of the available data. Similarly $m$  relates sample estimates, but we have instead simply defined it without the hat in Expression~\ref{mdef} to reduce clutter.} Similarly, any $m$ is to be evaluated/postulated from information outside the data. It parameterizes ``relative confounding'', comparing the strength of the $D\leftarrow Z \to Y$ relationship to the $D\leftarrow Z \to N$ relationship in Figure \ref{f_simple_dags}(a). This comparison thus comes down to $Z \to Y$ versus $Z \to N$.

\subsection{A scale-free transformation of $m$.}
\label{section_rexpress_bias_y}

One important concern with the parameterization above is that $N$ and $Y$ may be dissimilar enough in conception or measurement that we do not understand them to be on the same scale. This complicates the interpretation of $m$, which is sensitive to any scale differences between $N$ and $Y$. An analogous problem arises in sensitivitiy analysis with omitted variables on the ``coefficient'' scale as discussed in \cite{CinelliHazlett2020}, and we propose a similar solution: reparameterization in terms of ``(partial) variance explained'', i.e. partial-correlation parameters.

As a first intuition, one might expect that Equation~\ref{e_revised_estimate_y_trad_ovb_reexpress} could be rewritten in a form that rescales the bias adjustment according to the relative scales of $Y$ and $N$. To arrive at such a statement formally, let us first consider the case where we have a single omitted $Z$. \cite{CinelliHazlett2020} show that omitted variable bias can be written in terms of (i) the partial correlation between the outcome and $Z$ after accounting for $D$ and $X$ ($R_{Y \sim Z | D,X}$), (ii) (a transformation of) the partial correlation of the treatment and $Z$ after accounting for $X$ ($R_{D \sim Z |X}$), and (iii) a factor that accounts for the scales of the outcome and treatment. Specifically, 
\vspace{-.2in}
\begin{align} \label{e_cinelli_hazlett_single_z_y}
\widehat{\text{bias}}_{\text{(YD.X)}}  = \hat{\beta}_{Y\sim D | X} - \hat{\beta}_{Y\sim D| Z, X} &= \hat{R}_{Y \sim Z | D,X}  \frac{\hat{R}_{D\sim  Z |X}}{\sqrt{1-\hat{R}^2_{D\sim  Z |X}}} \frac{\widehat{\sd}(Y^{\perp D,X)}}{\widehat{\sd}(D^{\perp X})}\nonumber \\
&= \hat{R}_{Y \sim Z | D,X}  \hat{f}_{D\sim  Z |X} \frac{\widehat{\sd}(Y^{\perp D,X)}}{\widehat{\sd}(D^{\perp X})}
\end{align}

\noindent where we employ Cohen's $f$, i.e. $f_{D\sim  Z |X} = \frac{R_{D\sim  Z|X}}{\sqrt{1-R_{D\sim  Z|X}^2}}$, to simplify the second term.  Similarly, 
\vspace{-.2in}
\begin{equation} \label{e_cinelli_hazlett_single_z_n}
\widehat{\text{bias}}_{\text{(ND.X)}}  = \hat{\beta}_{N\sim D | X} - \hat{\beta}_{N\sim D| Z, X} = \hat{R}_{N \sim Z | D,X}  \hat{f}_{D\sim  Z |X} \frac{\widehat{\sd}(N^{\perp D,X)}}{\widehat{\sd}(D^{\perp X})}
\end{equation}

Because the bias contributed by $\hat{f}_{D \sim Z |X}$ is the same for both, it cancels out in the ratio of biases, $m$, which is now seen to be:
\vspace{-.2in}
\begin{equation} \label{e_m_to_k_single_z}
\begin{aligned}
m
= \frac{{\widehat{\text{bias}}_{\text{(YD.X)}}}}{{\widehat{\text{bias}}_{\text{(ND.X)}}}}
=  \underbrace{\frac{\hat{R}_{Y \sim Z | D,X}}{\hat{R}_{N \sim Z | D,X}}}_{k} \times \underbrace{\frac{\widehat{\sd}(Y^{\perp D,X})}{\widehat{\sd}(N^{\perp D,X})} }_{\text{SF}}
\end{aligned}
\end{equation}

In this arrangement, we have defined $k\overset{\Delta}{=} \frac{\hat{R}_{Y \sim Z | D,X}}{\hat{R}_{N \sim Z | D,X}}$, which compares the correlation resulting from $Z\to Y$ to the correlation resulting from $Z\to N$. Thus, rather than reason about $m$, we can can reason about the relative influence $Z$ has on $Y$ versus $N$. We also defined $\text{SF} \overset{\Delta}{=} \hat{\sd}(Y^{\perp D,X})/\hat{\sd}(N^{\perp D,X})$ as the ``scale factor'' that makes up for scale differences between $Y$ and $N$, after removing the influence of $D$ and $X$ (the numerator and denominator are standard deviations of the residuals from the short regressions, so this is estimable). Solving Equation~\ref{e_m_to_k_single_z} for $k$, we see that $k = m/\text{SF}$; thus $k$ can also be reasoned about similarly to $m$ -- the ratio of biases felt by $N$ and $Y$, rescaled by $\text{SF}$ to make up for scale differences between $Y$ and $N$.

In many settings, we may worry that there are potentially many omitted variables. When this is the case, we can consider omitted variable bias for $Y$ where $Z$ is replaced by $Z_{\text{(Y.DX)}} \overset{\Delta}{=} \mathbf{Z}\beta_{Y \sim Z|D,X}$, the linear combination of $\mathbf{Z}$ that ``de-confounds'' the $D\to Y$ relationship in the same way that including $\mathbf{Z}$ in Equation \ref{e_long_reg_outcome_y} does. Similarly, we can consider omitted variable bias for $N$ where $Z$ is replaced by $Z_{\text{(N.DX)}} \overset{\Delta}{=} \mathbf{Z}\beta_{N \sim Z|D,X}$, the linear combination of $\mathbf{Z}$ that ``de-confounds'' the $D\to N$ relationship in the same way that including $\mathbf{Z}$ in Equation \ref{e_long_reg_outcome_n} does. Importantly, $Z_{\text{(Y.DX)}}$ may not equal $Z_{\text{(N.DX)}}$ and so 
\vspace{-.2in}
\begin{equation}\label{kasratiobf}
k \overset{\Delta}{=} \frac{\hat{R}_{Y \sim Z_{\text{(Y.DX)}} | D,X}  \hat{f}_{D\sim  Z_{\text{(Y.DX)}}|X}}{\hat{R}_{N \sim Z_{\text{(N.DX)}} | D,X}  \hat{f}_{D\sim  Z_{\text{(N.DX)}}|X}}
\end{equation}
\noindent where both the numerator and the denominator are the bias due to omission of $Z$  with respect to the relevant regressions (see e.g. \ref{e_cinelli_hazlett_single_z_y}), up to a shared rescaling factor. While this is more complicated than the expressions in the single $Z$ case, it remains that $k$ will behave according to $k = m/\text{SF}$. It can thus be reasoned about similarly to $m$: the ratio of biases felt by $N$ and $Y$, but now rescaled to make  up for scale differences between $Y$ and $N$, after removing the influence of $D$ and $X$. 

Reasoning about $k$ can also be clarified by considering potential outcomes, since this points more directly to the biases of concern. In Expression~\ref{kasratiobf}, consider the outcome absent treatment, $Y(0)$ (though we could have chosen $Y(d)$ for either $d=0$ or $d=1$ or some other value if $D$ is not binary). The numerator can then be simplified as $\hat{R}_{Y(0) \sim Z_{\text{(Y(0).DX)}} | D,X}  \hat{f}_{D\sim  Z_{\text{(Y(0).DX)}}|X} = \hat{f}_{Y(0)\sim D | X}$ (see Appendix \ref{reexpress_BF}). This is simply (a transformation of) the sample correlation between $D$ and $Y(0)$ given $X$. That correlation is a measure of bias; it describes how strongly $Y$ will be associated with $D$ other than via a causal effect or observed covariates (i.e., the $D\leftarrow Z \to Y$ path). 
The same is true of the placebo outcome: writing $N(0)$ for the placebo outcome absent treatment, the denominator of Expression~\ref{kasratiobf} reduces to $\hat{f}_{N(0)\sim D | X}$, which is again a transformation of the correlation between $N$ and $D$ given $X$ even absent treatment, and thus a measure of bias. Altogether, these interpretations merely state equivalent ways of understanding $k$ as the ratio of biases, on the correlation scale, suffered by $D\to Y$ vs. that suffered by $D \to N$, due to the omission of $Z$.

\paragraph{Adjusting for bias using $K$.}Using $k$ rather than $m$, the bias and adjusted coefficient can simply be rewritten as
\vspace{-.2in}
\begin{equation} \label{e_bias_po}
\begin{aligned}
\widehat{\text{bias}}_{\text{(YD.X)}}
= k \times \left(\hat{\beta}_{N\sim D | X} - \hat{\beta}_{N\sim D| Z, X} \right) \times \frac{\widehat{\sd}(Y^{\perp D,X})}{\widehat{\sd}(N^{\perp D,X})}
\end{aligned}
\end{equation}
This bring us to the expression we will employ,\footnote{It is also possible to show that $\frac{\widehat{\sd}(Y^{\perp D,X})}{\widehat{\sd}(N^{\perp D,X})}= \frac{\left[ \frac{\widehat{\sd}(Y^{\perp D,X})}{\widehat{\sd}(D^{\perp X})} \right] }{ \left[\frac{\widehat{\sd}(N^{\perp D,X})}{\widehat{\sd}(D^{\perp X})}  \right] } =\frac{\widehat{\text{se}}(\hat\beta_{Y\sim D|X})}{\widehat{\text{se}}(\hat\beta_{N\sim D|X})} \times \sqrt{\frac{\text{df}_Y }{\text{df}_N }}$, which is an expression containing only commonly reported regression summary statistics from the two short regressions that are estimable from the observed data. 
} 
\vspace{-.2in}
\begin{equation} \label{e_revised_estimate_po}
\begin{aligned}
\hat{\beta}_{Y\sim D| Z, X} &= \hat{\beta}_{Y\sim D | X} - k  \times \left(\hat{\beta}_{N\sim D | X} - \hat{\beta}_{N\sim D| Z, X} \right) \times \frac{\widehat{\sd}(Y^{\perp D,X})}{\widehat{\sd}(N^{\perp D,X})}
\end{aligned}
\end{equation}

Given $k$ and $\hat{\beta}_{N\sim D| Z, X}$, the estimator $\hat{\beta}_{Y\sim D| Z, X}$ from Equation \ref{e_revised_estimate_po} is consistent for $\hat{\beta}_{Y\sim D| Z, X}$ and asymptotically normal. Inference can be conducted using the non-parameteric bootstrap, or by constructing bootstrap standard errors and applying the normal approximation. 

\subsection{Uses of partial identification in practice.} These results are not merely of theoretical value but rather aim to solve a problem faced by investigators making real-world inferences. Similar to other sensitivity analysis and partial identification settings, we consider to main modes of inference. The first approach is ``postulative partial identification'': one reasons about the imperfection in the placebo (decide how far $\hat{\beta}_{N\sim D| Z, X}$ could arguably be from zero), and considers $k$ (the ratio of confounding strengths experienced by $D\to N$ versus $D\to Y$). Users are unlikely to be able to defend point values on these parameters, but rather consider plausible ranges, and thus boundaries beyond which the values are arguably absurd. For all such values in the plausible range, the adjusted coefficients for $\hat\beta_{Y\sim D| Z, X}$ can be computed.  

The alternative, isomorphic, inferential approach is the inverse procedure: look for the boundaries where the adjusted estimate substantively changes and then ask if the assumptions at those boundaries can be defended, cannot be defended but are plausible, or are arguably implausible. We illustrate these approaches in simulations and real data applications in Section~\ref{sec:applications}.


\subsection{Placebo treatments}
\label{proposal_pt}

Similar machinery can leverage imperfect placebo treatments rather than imperfect placebo outcomes. Suppose that we want to run the long regression for $Y$ on $D$ (Equation \ref{e_long_reg_outcome_y}) but, again, $\mathbf{Z}$ is an unobserved vector of variables that we would have liked to include in the regression but cannot. Suppose also that we have a placebo treatment, $P$, for which we believe the confounding of $P$ and $Y$ from $\mathbf{Z}$ can be compared to the confounding of $D$ and $Y$ from $\mathbf{Z}$.
We might consider running two separate regressions as we did for placebo outcomes. However, that approach is not easily adapted to imperfect placebo treatments, where there may be a direct relationship between $P$ and $Y$.\footnote{Suppose we have a causal model like Figure \ref{f_simple_dags}(d). $\hat{\beta}_{Y \sim Z|P,X}$ (or $\hat{R}_{Y \sim Z|P,X}$) will capture the causal relationship $D\to Y$, since the path $Z\to D \to Y$ will remain open conditional on $P$ and $X$ alone. Expressing the bias in the regression coefficient on $D$ from the short regression using a two regression approach like we used for placebo outcomes would, therefore, require us to reason about the causal effect of interest itself, which is not useful.}
Instead, suppose that $P$ is not a descendant of $D$, $D$ is not a descendant of $P$, $P$ is a good or neutral control for the effect of $D$ on $Y$, and $P$ is a good or neutral control for $D$ on $Y$, conditional on $\mathbf{Z}$ and $\mathbf{X}$.\footnote{We require that $D$ and $P$ are good or neutral controls for each other in that they do not increase the bias in our estimate of the coefficient we are targeting ($\hat{\beta}_{Y\sim D|X,Z}$).
This will simplify interpretation. If a user is concerned that there could be a path opened or closed by conditioning on $D$ or $P$, we suggest drawing the DAG to check. See the Supplement for further discussion.} 
Further, suppose that we want to run the ``long'' regression of $Y$ on $D,P,\mathbf{X}$, and $\mathbf{Z}$ in Equation \ref{e_long_reg_treatment_y_mt}. In particular, we are interested in the coefficient $\beta_{Y\sim D| P, Z, X}$. However, since $\mathbf{Z}$ is unobserved, we must run the ``short'' regression of $Y$ on $D,P$ and $\mathbf{X}$ in Equation \ref{e_short_reg_treatment_y_mt}, rather than the desired long regression,
\vspace{-.2in}
\begin{equation} \label{e_long_reg_treatment_y_mt}
\begin{aligned}
Y &= \beta_{Y\sim D| P, Z, X} D + \beta_{Y\sim P| D, Z, X} P + \mathbf{X}\beta_{Y \sim X| D,P,Z} +  \mathbf{Z}\beta_{Y \sim Z|D,P,X}  + \epsilon_l
\end{aligned}
\end{equation}
\begin{equation} \label{e_short_reg_treatment_y_mt}
\begin{aligned}
Y &= \beta_{Y\sim D | P,X} D + \beta_{Y\sim P| D, X} P + \mathbf{X}\beta_{Y \sim X| D,P} + \epsilon_s
\end{aligned}
\end{equation}

\noindent Following a similar approach as we took for placebo outcomes, we can arrive at similar partial identification expressions for the placebo treatment case,
\vspace{-.2in}
\begin{equation} \label{e_bias_pt_mt}
\begin{aligned}
\widehat{\text{bias}}_{\text{(YD.PX)}} \overset{\Delta}{=} \hat{\beta}_{Y\sim D | P,X} - \hat{\beta}_{Y\sim D| P, Z, X} 
=  k \times \left(\hat{\beta}_{Y\sim P | D,X} - \hat{\beta}_{Y\sim P| D,Z, X}\right) \times \frac{\widehat{\sd}(P^{\perp D,X})}{\widehat{\sd}(D^{\perp P,X})}
\end{aligned}
\end{equation}
\begin{equation} \label{e_revised_estimate_pt_mt}
\begin{aligned}
\hat{\beta}_{Y\sim D| P, Z, X}
= \hat{\beta}_{Y\sim D | P,X} - k \times \left(\hat{\beta}_{Y\sim P | D,X} - \hat{\beta}_{Y\sim P| D,Z, X}\right) \times \frac{\widehat{\sd}(P^{\perp D,X})}{\widehat{\sd}(D^{\perp P,X})}
\end{aligned}
\end{equation}

All components of these expressions except for the two sensitivity parameters can be estimated from the data.\footnote{It is also possible to show that 
$\frac{\widehat{\sd}(P^{\perp D,X})}{\widehat{\sd}(D^{\perp P,X})}
= \frac{\widehat{\sd}(P^{\perp D,X})}{\widehat{\sd}(D^{\perp P,X})}
\times \frac{\widehat{\sd}(Y^{\perp D,P,X})}{\widehat{\sd}(Y^{\perp D,P,X})} =\frac{\widehat{\se}(\hat\beta_{Y\sim D|P,X})}{\widehat{se}(\hat\beta_{Y\sim P|D,X})} \times \sqrt{\frac{\text{df}_D }{\text{df}_P }}$
which requires only commonly reported regression summary statistics from the short regression.
} $\hat{\beta}_{Y\sim P| D,Z, X}$ parameterizes any direct causal relationship between $P$ and $Y$ (placebo imperfection). 
As with placebo outcomes, $k$ captures the relative levels of confounding. Similar to the placebo outcome case, the $\mathbf{Z}$ to contemplate here is that which contains a rich enough set of unobserved variables (and no problematic variables\footnote{Note that since we can extend $\mathbf{Z}$ arbitrarily, we actually do not need to worry about conditioning on pre-treatment colliders in the same way as we would when conditioning on observed variables. Suppose we have the canonical M-shaped DAG. If the central collider node is included in $\mathbf{Z}$ then we can also simply include one of its parents that create the collider to block the M-shaped backdoor path. Taken further, we could consider all variables along backdoor paths as included in $\mathbf{Z}$, if necessary.}) to de-confound both $P\to Y$ and $D\to Y$. Note also that, because we consider a single regression with $Y$ as the outcome that includes both $D$ and $P$ on the right-hand side,  we can compare the level of confounding of the $P\to Y$ relationship from $\mathbf{Z}$ to the level of confounding of the $D\to Y$ relationship from $\mathbf{Z}$ by using the same linear combination, $Z_{\text{(Y.DPX)}} = \mathbf{Z}\beta_{Y \sim Z|D,P,X}$. 

Here, $k$ is the ratio of the level of total confounding of $D\to Y$ from $D\leftarrow Z \to Y$ to the level of total confounding of $P\to Y$ from $P\leftarrow Z \to Y$, after re-scaling one of these biases to account for scale differences between $P$ and $D$. This defines $k$ as $k \overset{\Delta}{=} \frac{\hat{R}_{Y\sim Z_{\text{(Y.DPX)}}|D,P,X}\hat{f}_{D\sim Z_{\text{(Y.DPX)}}|P,X}}{\hat{R}_{Y\sim Z_{\text{(Y.DPX)}}|D,P,X}\hat{f}_{P\sim Z_{\text{(Y.DPX)}}|D,X}} = \frac{\widehat{\text{bias}}_{\text{(YD.PX)}}}{\widehat{\text{bias}}_{\text{(YP.DX)}}\times \text{SF}}$. 
In the special case when there is only a single omitted variable, then $k$ can also be written as $k = \frac{\hat{f}_{D\sim Z|P,X}}{\hat{f}_{P\sim Z|D,X}}$. This can be thought of as comparing the strength of the $Z\to D$ ``arm'' of the confounding path $D\leftarrow Z \to Y$ to the relative strength of the $Z\to P$ ``arm'' of the confounding path $P\leftarrow Z \to Y$ in Figure \ref{f_simple_dags}(b).

\subsection{Varieties of (single) placebos}
\label{taxonomy}

The approaches presented above are algebraic properties of linear regression, and apply to any pair of regressions. However, an assumed causal structure (embodied by a graph) is necessary to link those results to causal quantities of interest. While Figure~\ref{f_simple_dags} represents four graphs in which a placebo variable can be considered a placebo outcome or treatment, there are additional cases representing placebos as mediators, observed confounders, or post-outcome variables. Tables~\ref{taxonomy_table1} and ~\ref{taxonomy_table2} illustrate 8 different causal structure with a single placebo variable in different positions, inclusive of those in Figure~\ref{f_simple_dags}. We assume that all relevant unobserved common causes are bundled into the vector of unobserved variables $\mathbf{Z}$; and we refer to the placebo as $P$ in every case as the distinction between placebo treatments and placebo outcomes 
is not always appropriate or clear. We omit observed covariates $\mathbf{X}$ from the graphs for legibility. These tables detail the specific regressions one would run and adjustment one would make for partial identification purposes in each case. In some causal structures, a given placebo variable can be understood to have different roles, allowing it to be handled in more than one way as separate exercises that may lead to informative triangulation.

Consider Table \ref{taxonomy_table1}[c]. The causal structure includes the edge $P\to Y$. There are two options for how we might approach this setting. First, we might consider $P$ to be an \textit{imperfect} placebo treatment. The machinery that we detailed above (and outlined in the entry for Table \ref{taxonomy_table1}[a] - Placebo Treatment) is then sufficient. Second, we might consider $P$ to be a placebo outcome. Suppose we simply applied the placebo outcome approach we've already seen. In this case, $\text{bias}_{\text{(YD.X)}}$, the numerator of $k$, will capture both $D\leftarrow Z \to Y$ and $D\leftarrow Z \to P \to Y$, while $\text{bias}_{\text{(PD.X)}}$, the denominator of $k$, will capture $D\leftarrow Z\to P$. $D\leftarrow Z\to P$ appears in both the numerator and denominator of $k$, making the parameter more difficult to interpret. The solution is straightforward: include $P$ in the regression where $Y$ is the outcome. We see that the ``Short Regressions'' column of the table shows this update. $P$ plays a role like an observed confounder: Controlling for it blocks a non-causal path between $D$ and $Y$.  In Appendix \ref{SPDs}, we derive the expression for $\hat{\beta}_{Y\sim D|P,Z,X}$ that we see in the ``Target Coefficient Expression'' column. This contains two coefficients from the short regressions and a scale factor that are all estimable from observed data (i.e., data on $Y,D,P,X$). It also contains a $k$ parameter that compares the level of unobserved confounding from $D\leftarrow Z \to Y$ (i.e., $\widehat{\text{bias}}_{\text{(YD.PX)}}$) to that from $D\leftarrow Z\to P$ (i.e., $\widehat{\text{bias}}_{\text{(PD.X)}}$), after accounting for scale differences, and an imperfect placebo parameter capturing the direct causal relationship between $D$ and $P$, which in this graph is zero. These are listed in the ``Parameters'' column. The remaining entries in Tables \ref{taxonomy_table1} and \ref{taxonomy_table2} can be read similarly. When the causal structure is that of a mediator (i.e., there exists a path $D\to P\to Y$ in addition to $D\to Y$), the placebo approach becomes very complicated, and we do not recommend our approach.

\subsection{Relaxed difference in difference as a special case}
\label{section_DID}

In this section, we discuss how traditional DID is a special case of the placebo outcome approach. This connection helps to illustrate the narrowness of the parallel trends assumption, and shows how the placebo outcome framework offers a relaxation or sensitivity analysis that can be used to arrive at more defensible bounds on treatment effects. 

To begin with a brief and intuitive explanation for the connection, consider  Expression~\ref{e_revised_estimate_y_trad_ovb_reexpress} above, but removing covariates $X$ for simplicity:
\vspace{-.2in}
\begin{equation}
\hat{\beta}_{Y\sim D| Z} = \hat{\beta}_{Y\sim D} - m \times \left(\hat{\beta}_{P\sim D} - \hat{\beta}_{P\sim D| Z} \right)
\end{equation}
\noindent where $Y$ is the outcome in the post-treatment period and $P$ is the outcome in the pre-treatment period, acting here as a placebo outcome. The first term on the right hand side, $\hat{\beta}_{Y\sim D}$, is in fact the difference in means (DIM) on the (post-treatment) outcome, comparing the treated group to the control group. The DIM comparing treated and control groups on the pre-treatment outcome is likewise given by $\hat{\beta}_{P\sim D}$. For a perfect placebo outcome the final term ($\hat{\beta}_{P\sim D| Z}$) would equal zero in expectation. This is arguably the case in the DID setting, because the placebo outcome is a pre-treatment measure and so cannot be affected by treatment. Finally, to argue that $m=1$ would be to argue that the bias in group difference on the post-treatment outcome is exactly equal to the group difference on the pre-treatment outcome. This is precisely the ``parallel trends'' assumption. Putting this together, assuming a perfect placebo outcome in the sample ($\hat{\beta}_{P\sim D| Z}=0$) and parallel trends ($m=1$), we are left with  $\hat{\beta}_{Y\sim D| Z} = \hat{\beta}_{Y\sim D} - \hat{\beta}_{P\sim D}$, which is identical to the DID estimate.

While the above connection to DID estimation is useful, we can also explicitly connect this with the ATT estimand using more formal causal logic as follows. On the level of estimands, consider the following decompositions of the difference in means (DIM) estimand for binary treatments. For an outcome $Y$ and binary treatment where the treated group is indicated by $G$, 
\vspace{-.2in}
\begin{equation} \label{e_did_att_decomp_y}
\begin{aligned}
\text{DIM}_Y
&= \mathbb{E}[Y|G=1] - \mathbb{E}[Y|G=0] 
= \mathbb{E}[Y_1|G=1] - \mathbb{E}[Y_0|G=0] \\
&= \underbrace{\mathbb{E}[Y_1|G=1] - \mathbb{E}[Y_0|G=1]}_{\text{ATT}_Y} + \underbrace{\mathbb{E}[Y_0|G=1] - \mathbb{E}[Y_0|G=0]}_{\text{Bias}_Y} 
\end{aligned}
\end{equation}

\noindent Similarly, for a placebo outcome $P$,  
\vspace{-.2in}
\begin{equation} \label{e_did_att_decomp_n}
\begin{aligned}
\text{DIM}_P
&= \mathbb{E}[P|G=1] - \mathbb{E}[P|G=0] 
= \mathbb{E}[P_1|G=1] - \mathbb{E}[P_0|G=0] \\
&= \underbrace{\mathbb{E}[P_1|G=1] - \mathbb{E}[P_0|G=1]}_{\text{ATT}_P} + \underbrace{\mathbb{E}[P_0|G=1] - \mathbb{E}[P_0|G=0]}_{\text{Bias}_P} 
\end{aligned}
\end{equation}

\noindent With these in hand, it is easy to see that there exists an $m\in \mathbb{R}$ such that we can obtain the expression for $\text{ATT}_Y$,
\vspace{-.2in}
\begin{equation} \label{e_did_att_expression}
\begin{aligned}
\text{ATT}_Y
&= \text{DIM}_Y - \text{Bias}_Y \\
&= \text{DIM}_Y - m \times \text{Bias}_P \\
&= \text{DIM}_Y - m \times (\text{DIM}_P - \text{ATT}_P), \text{ where } m = \frac{\text{Bias}_Y}{\text{Bias}_P}
\end{aligned}
\end{equation}

Because $P$ represents a pre-treatment outcome, it is often unproblematic to assume that $\text{ATT}_P = 0$. The more challenging assumption of DID, however, is ``parallel trends'': the expected overtime change in non-treatment outcome in the treated group and control group are identical. It is easy to show that the parallel trends assumption is equivalent to an assumption of equiconfounding ($m=1$) (Equation \ref{e_did_pt_is_ec_2}; \cite{Sofer2016}).  
\vspace{-.2in}
\begin{equation} \label{e_did_pt_is_ec_2}
\begin{aligned}
\text{Parallel Trends} \overset{\Delta}{=}
\underbrace{\mathbb{E}[Y_0|G=1] - \mathbb{E}[P_0|G=1]}_{\text{Trend}_{G=1}} &= \underbrace{\mathbb{E}[Y_0|G=0] - \mathbb{E}[P_0|G=0]}_{\text{Trend}_{G=0}} \\
\iff 
\underbrace{\mathbb{E}[Y_0|G=1] - \mathbb{E}[Y_0|G=0]}_{\text{Bias}_Y} &= \underbrace{\mathbb{E}[P_0|G=1] - \mathbb{E}[P_0|G=0]}_{\text{Bias}_P}
\overset{\Delta}{=} \text{Equiconfounding} 
\end{aligned}
\end{equation}

To connect this with the approach above: Since $G$ is binary, a regression of $Y$ (or $P$) on $G$ will give the difference in means $\text{DIM}_Y$ (or $\text{DIM}_P$), respectively, as the coefficient on $G$. Under the standard DID assumptions, we would have $\text{ATT}_Y = \beta_{Y\sim G}- \beta_{P\sim G}$. To relax parallel trends/ equiconfounding, using omitted variable bias, we instead consider

\vspace{-.2in}
\begin{equation} \label{e_did_NP_to_P_connection_y}
\begin{aligned}
\text{ATT}_Y + \text{Bias}_Y = \text{DIM}_Y = \beta_{Y\sim G} = \beta_{Y\sim G|Z} + \text{Bias}_{Y,\text{reg.}}
\end{aligned}
\end{equation}
\begin{equation} \label{e_did_NP_to_P_connection_n}
\begin{aligned}
\text{ATT}_P + \text{Bias}_P = \text{DIM}_P = \beta_{P\sim G} = \beta_{P\sim G|Z} + \text{Bias}_{P,\text{reg.}}
\end{aligned}
\end{equation}

\noindent where $\mathbf{Z}$ contains all the remaining confounders of the $G,Y$ and $G,N$ relationships. Under assumptions of effect homogeneity, or using OLS regression as an approximation, we have that $\text{ATT}_Y = \beta_{Y\sim G|Z}$, which means that $\text{Bias}_Y = \text{Bias}_{Y,\text{reg.}}$. Similarly, $\text{ATT}_P = \beta_{P\sim G|Z}$, which means that $\text{Bias}_P = \text{Bias}_{P,\text{reg.}}$.
Plugging into the expression for $\text{ATT}_Y$, we get Equation \ref{e_did_reg_id}, where $m$ is the ratio of biases for the regression coefficients.
\vspace{-.2in}
\begin{equation} \label{e_did_reg_id}
\begin{aligned}
\text{ATT}_Y 
&= \beta_{Y\sim G} - m\times (\beta_{P\sim G} - \beta_{P\sim G|Z})
\end{aligned}
\end{equation}

As in previous sections, we could re-express $m$ to account for scale differences.  If desired, assumptions on $m$ can be mapped to assumptions on differential trends (see Appendix~\ref{app_m_to_w}).

\section{Applications}\label{sec:applications}

\subsection{Simulation evidence with a placebo treatment}

Before applying the approach, we demonstrate it and verify that it behaves as expected in a simple  simulation representing a plausible placebo treatment example. Suppose we are interested in the effect that physical activity has on cardiovascular health for adults in the United States. We have a representative sample of adults with data on whether they exercise regularly or not ($D$), whether they take supplemental vitamins ($P$), and whether they experience heart disease over a fixed follow up period ($Y$). We may worry that socioeconomic status ($Z$), a variable we do not observe in our data, confounds the observed exercise--heart disease ($D\to Y$) relationship. For example, socioeconomic status could influence whether or not individuals are able to exercise regularly while influencing risk of heart disease through numerous other channels (diet, access to healthcare, quality of living conditions, stress, etc). We also suspect that such factors could influence whether or not individuals demonstrate other health seeking behaviors, such as supplemental vitamins. Further, suppose that we expect these supplements to have little or no effect on heart disease. Thus,  supplemental vitamins ($P$) can be used as a (possibly imperfect) placebo treatment in our effort to determine the causal effect of regular exercise on heart disease.

To simulate data, we use the following data generating process. 
We generate a sample of 5000 adults. Socioeconomic status ($Z$) is a latent variable, for which we use a uniform random variable over $[0,1]$. Whether or not someone does some amount of physical exercise ($D$) is a Bernoulli random variable where the probability of physical exercise is $\Phi(2 Z)$, where $\Phi$ is the CDF for the standard normal distribution. Higher socioeconomic status increases the probability of doing physical exercise. This leads to $Pr(D=1)$ (the proportion doing some exercise) of roughly 0.8. Whether or not someone takes supplemental vitamins ($P$) is a Bernoulli random variable equal to 1 with probability $\Phi(-1 + Z)$, so that higher socioeconomic status also increases the probability of taking supplemental vitamins. The resulting $Pr(P=1)$ is roughly 30\%. Finally for the outcome, whether or not someone experiences heart disease ($Y$), we use a Bernoulli draw with probability $\Phi(-2 - 0.43 D + 4 Z)$. About 40\% of the sample has heart disease ($Y=1$). 
The target quantity is the sample estimate one would get for the coefficient on exercise, i.e. $\hat{\beta}_{Y\sim D| P, Z}$ from the linear probability model  $Y = \beta_{Y\sim D| P, Z} D + \beta_{Y\sim P| D, Z} P +  \beta_{Y \sim Z|D,P}Z  + \epsilon_l$ if it could be fitted in the observed sample. The value of $\hat{\beta}_{Y\sim D| P, Z}$ in each simulated dataset is roughly -0.1, i.e. controlling for socioeconomic status and taking supplemental vitamins, the individuals who do some physical activity have 10\% lower probability of having heart disease. 

Returning to the perspective of a researcher given these data, we target the infeasible ``long'' regression, $Y = \beta_{Y\sim D| P, Z} D + \beta_{Y\sim P| D, Z} P +  \beta_{Y \sim Z|D,P}Z  + \epsilon_l$ where $\beta_{Y\sim D| P, Z}$ is of interest. However, we do not observe socioeconomic status ($Z$) and so instead can only estimate the ``short'' regression $Y = \beta_{Y\sim D | P} D + \beta_{Y\sim P| D} P  + \epsilon_s$. Following the approach above, we can partially identify the coefficient estimate in the sample, $\hat{\beta}_{Y\sim D| P, Z}$ with
\vspace{-.2in}
\begin{equation}
\hat{\beta}_{Y\sim D| P, Z}
= \hat{\beta}_{Y\sim D | P} - k \times \left(\hat{\beta}_{Y\sim P | D} - \hat{\beta}_{Y\sim P| D,Z}\right) \times \text{SF}
\end{equation}
\noindent where $k = \frac{\hat{f}_{D\sim Z|P}}{\hat{f}_{P\sim Z|D}} = \frac{\widehat{\text{bias}}_{\text{(YD.P)}}}{\widehat{\text{bias}}_{\text{(YP.D)}}\times \text{SF}}$ 
compares the correlation from $Z\to D$ to that from $Z\to P$ or, equivalently, the bias due to  $D\leftarrow Z\to Y$ to that from $P \leftarrow Z\to Y$. $\hat{\beta}_{Y\sim P| D,Z}$ captures any direct causal relationship between $P$ and $Y$. This would be $0$ if the placebo treatment ($P$, vitamins) is indeed ineffectual against heart disease. 
Finally $\text{SF}= \frac{\widehat{\sd}(P^{\perp D})}{\widehat{\sd}(D^{\perp P})}$ and $\hat{\beta}_{Y\sim P| D}$ can be estimated from the data.

\begin{figure}[h!]
\begin{center}
\caption{\label{f_HDSim_ContourPlot} Contour plot for physical activity and heart disease simulation.}
\includegraphics[scale=0.18]{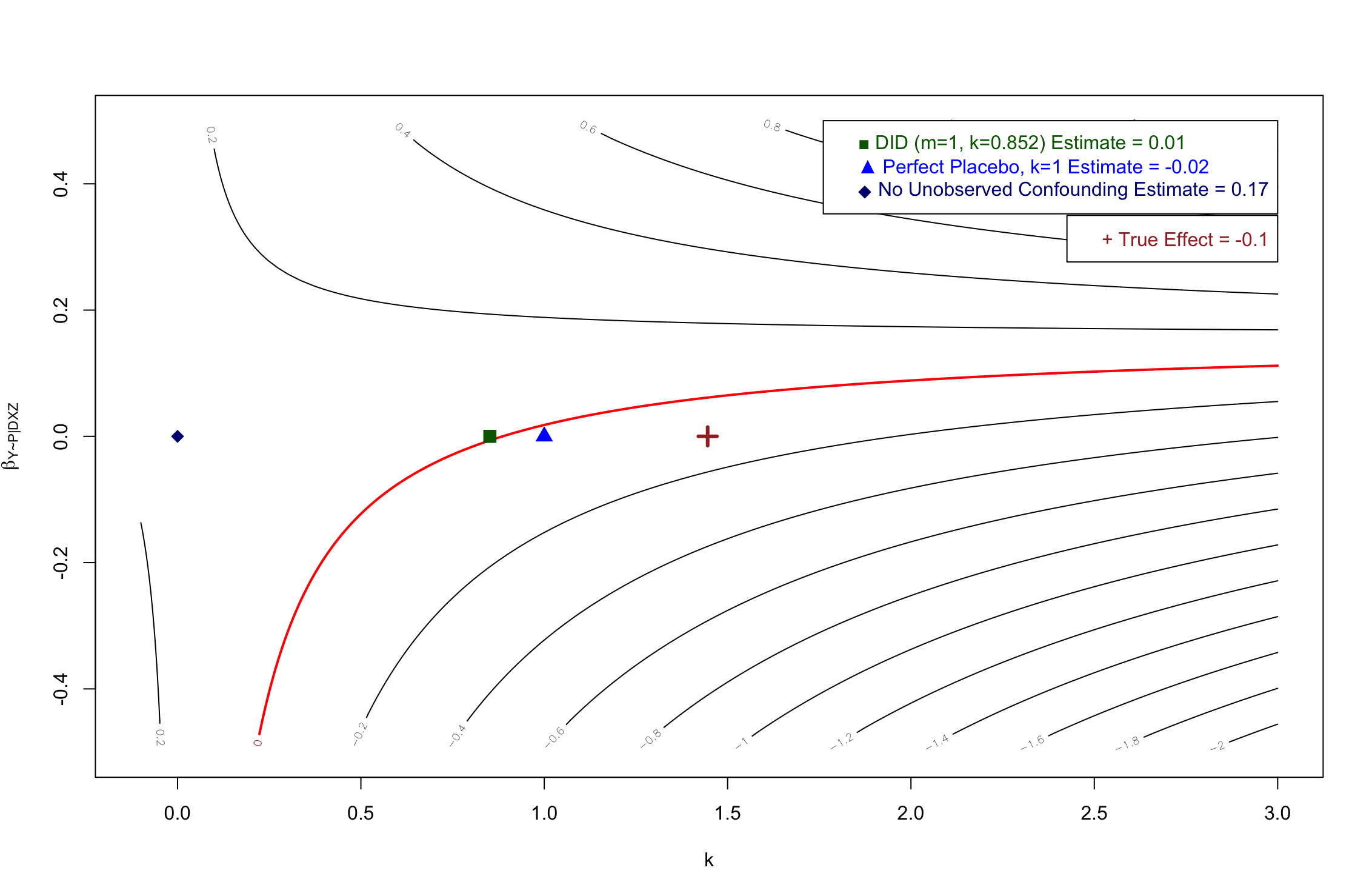}
\end{center}
\end{figure}

\begin{figure}[h!]
\begin{center}
\caption{\label{f_HDSim_LinePlot} Line plot for physical activity and heart disease simulation. $\beta_{Y\sim P| D,Z}=0$}
\includegraphics[scale=0.18]{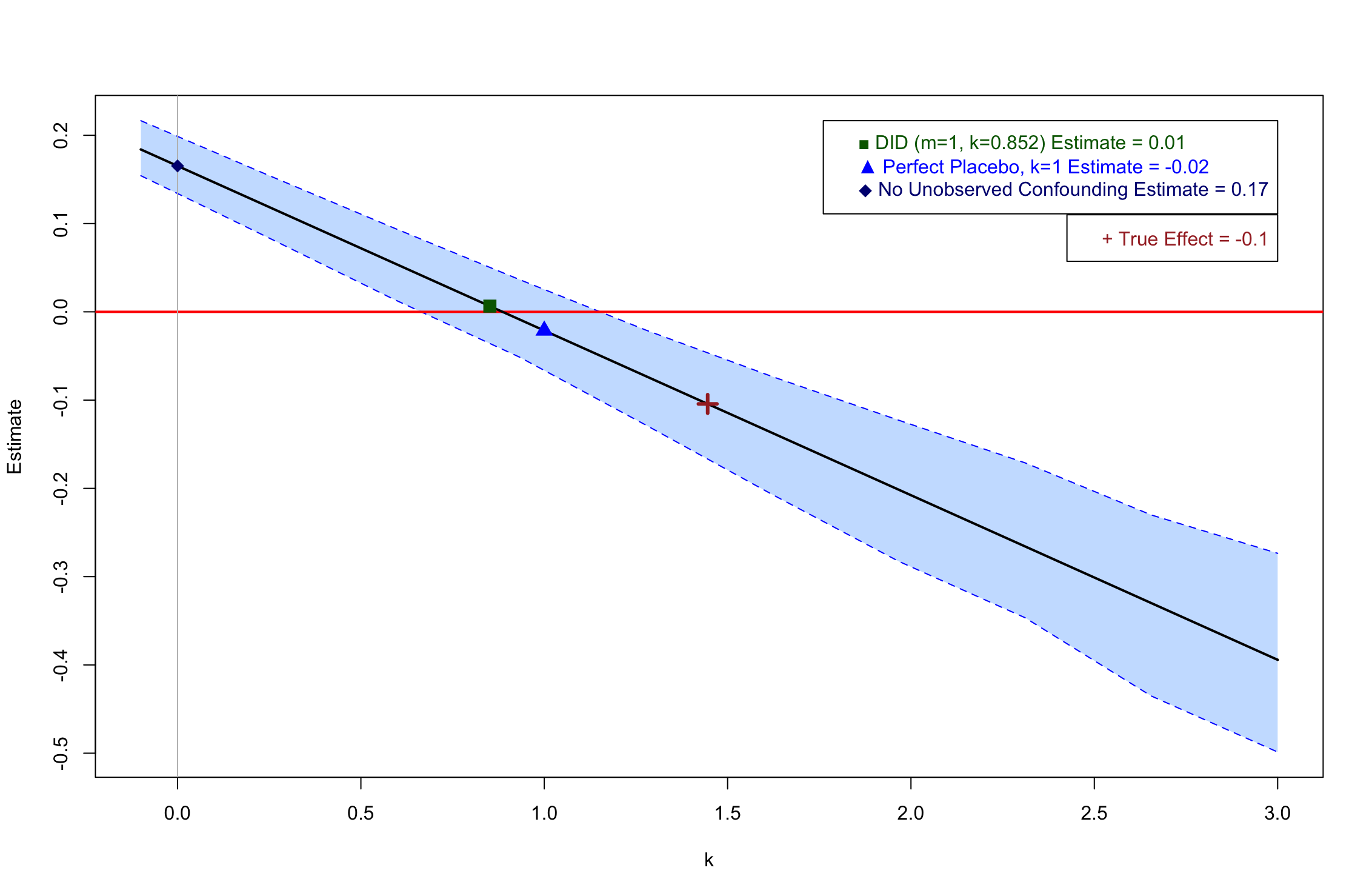}
\end{center}
\end{figure}

In practice, this analysis can be implemented easily in a few lines of code as shown in Appendix~\ref{code}.\footnote{We perform these analyses using the \texttt{placeboLM} 
package the authors built for this purpose, currently under development. It can be installed via \texttt{devtools::install\_github("Adam-Rohde/placeboLM")}. See \url{https://github.com/Adam-Rohde/PlaceboLM}.} As this is a simulation with data we generated, it is a fiction to reason about $k$ by reference to substantive real world knowledge. However, to illustrate how such reasoning would occur (and as the basis for the choice of simulation parameters) we mimic such a reasoning process briefly. First, in many real settings the researcher may be equipped to argue that the confounding bias suffered by the placebo-outcome relationship at least has the same sign as that suffered by the real treatment-outcome relations, and thus $k$ is non-negative. Second, to judge whether  $k>1$ is to ask if the true treatment (exercise) is ``more confounded'' with the outcome than the placebo treatment (vitamins). In this case, while we would not definitely rule-out $k<1$, we would most expect that $k>1$ because while some degree of vitamin supplementation can be achieved relatively cheaply and with little investment of time and effort, having the spare time, energy, and income to spend on regular exercise may depend more strongly on socioeconomic status. That said, 
both vitamin supplementation and regular exercise are likely somewhat dependent on socioeconomic status, so we may argue that while $k$ is certainly greater than 0 and probably greater than 1, it is perhaps not above 2, and quite unlikely to be above 3. Altogether then, a researcher in this setting might wish to see the results over the range $0 \leq k \leq 3$ to reflect the full set of estimates that cannot be strongly ruled out, while perhaps also placing the (\textit{ex ante}) gamble that $k$ is more likely between 1 and 2. We will see examples of reasoning about $k$ in real-world circumstances in the two applications below.

Returning to the estimation and interpretation process, we may first visualize the results most completely using a contour plot (Figure~\ref{f_HDSim_ContourPlot}) that shows the desired quantity ($\hat{\beta}_{Y\sim D| P,Z}$) across postulated values of both placebo imperfection ($\hat{\beta}_{Y\sim P| D, Z}$) and relative confounding ($k$). On this plot we show three point estimates and how they compare to the true estimate. First, the traditional strategy of simply assuming no unobserved confounding would have produced a poor estimate (+17pp). For the remaining two placebo-adjusted estimates, we assume precisely zero placebo imperfection ($\beta_{Y\sim P| D, Z}=0$). The conventional DID estimate $m=1$ corresponds to $k=0.85$ in this case, and produces an estimate of +1pp for the adjusted value of $\hat{\beta}_{Y\sim D| P, Z}$.  If we instead assume equal confounding after rescaling ($k=1$), we obtain an estimate of -2pp.  Both are much closer to the true effect estimate in the simulation of -10pp than assuming no unobserved confounding but neither are good approximations to the true effect. 

Next, it is likely easier for many readers to examine just one slice of Figure~\ref{f_HDSim_ContourPlot}, transformed into a two-dimensional line-plot now with the adjusted estimate shown on the vertical axis (Figure~\ref{f_HDSim_LinePlot}). Again we consider the range $0 \leq k \leq 3$ and overlay the point estimates that would be produced by assuming precisely zero unobserved confounding (and hence $k=0$), parallel trends on the original scale (DID, $m=1$), and equiconfounding in partial variance explained/ after rescaling ($k=1$). Note that in this particular simulation, the scale of $D$ and $P$ differ somewhat with $\text{SF} = 1.17$ and the true value for $k$ is 1.45. Had this been a real data example in which investigators must reason about the plausible values of $k$, then the correct answer would be contained so long as they argued $k=1.45$ was possible, i.e. that the confounding of the placebo treatment (vitamin use) with heart disease could be 1.45 times as severe as the confounding of the true treatment (regular exercise) with heart disease.

\subsection{Estimating the effect of a job training program on earnings}

The National Supported Work Demonstration (NSW) was a job training program. From 1975 to 1979, it aimed to help disadvantaged workers build basic skills. The program randomly assigned participants to training positions. These consisted of a group that received the benefits of the program (the treatment group) and a group that did not receive any benefits (the control group) \citep{LaLonde1986}. The data also include a non-experimental dataset in which individuals not from the NSW program but from the Panel Study of Income Dynamics (PSID) are added as additional control units. The data that we work with is a subset that consists of males with three (2 pre-training and 1 post-training) years of earnings data (the outcome of interest) constructed by \citep{dehejia1999causal}. 
We use the non-experimental dataset to illustrate the methods discussed in this paper and then compare these results with the estimates from the experimental data. The difference in means estimate of the treatment effect from the experimental data is 1,794 USD higher 1978 earnings in dollars for the treated group relative to the control group. The estimate is 1,671 USD when we additionally control for observed covariates. The experimental data are also very useful as they allow us to back out an estimate for the true value for $k$. 

Data on both 1974 earnings and 1975 earnings are available. We use 1975 earnings as a placebo outcome ($P$) for the real outcome, 1978 earnings ($Y$). Results are very similar when instead using 1974 earnings are the placebo (see below where we consider these results for triangulation and Appendix~\ref{code} for the full analysis with 1974 earnings). The treatment ($D$) is receiving the job training program. We also use the covariates ($X$) in the data including age, education (years and degree or not), marital status, and demographic indicators. 

Again the process is to first recognize that we would like to estimate the the ``long'' regression 
$Y = \beta_{Y\sim D| Z, X} D + \mathbf{X}\beta_{Y \sim X| D,Z} + \mathbf{Z}\beta_{Y \sim Z|D,X}   + \epsilon_l$, 
\noindent inclusive of $Z$ sufficient to handle confounding. Absent $Z$ we leverage 
the placebo outcome, and pursue partial identification or sensitivity analysis given postulated values of the relative confounding ($k$) and placebo imperfection ($\hat{\beta}_{P\sim D| Z, X}$). Since the placebo outcome is pre-treatment earnings, it is reasonable to assume that the treatment has no causal effect on the the placebo outcome, so we can comfortably assume that in expectation at least $\hat{\beta}_{P\sim D| Z, X}=0$.

How how would a researcher go about considering values for $k$?  First, we would expect the confounding bias experienced by $D\to Y$ to have the same sign as that experienced by $D\to P$, so $k>0$. Second, $P$ and $Y$ represent a person's income a few years apart, and we would expect many of the confounders of low-income with selecting into the program have some persistence over time: those likely to have low-income in 1975 and to therefore enroll in the study (compared to those in the PSID sample) will have some tendency to also have lower incomes in 1978. 
We might initially expect similar strength of confounding for $P$ and $Y$, such as $0.5 < k < 1.5$ as a starting point. However, because people likely select into the program based on current or very recent income/employment, we can except that confounding is stronger closer to the time of actual enrollment (1975) than years laters.  
Thus, we can reasonably expect $0 < k <1$. A researcher might further choose to place highest plausibility on the region $0.5<k<1$, as we do expect non-trivial confounding still present in the 1978 outcome.

On a historical note, the argument that people select into job training programs due to a recent loss of income is used here to suggest $k<1$ above, but appears in other forms of discussion such as regression to the mean broadly, or specifically in this case as the ``Ashenfelter dip'', which originated in relation to job training training programs of this kind \cite{ashenfelter1978estimating}. While the general observation that those who select into treatment might ``get better on their own'' serves to warn us of potential bias in either a within-subject design or DID, here it informs our choice of $k$.

We also note complications with the use of DID here. First, it assumes parallel trends on the data on some scale. observed scale. However, income changes in scale over time,  one might expect a change in the effective magnitude of bias, in dollars, from even an unchanging degree of confounding. Inflation alone in this period makes one US dollar in 1975 worth approximately 1.21 US dollars in 1978.
More fundamentally, an investigator may wish to use DID in cases like this as it appears to produce ``an answer''. This apparent certainty is, however, is a fiction. As subjective or unscientific is may seem to require ``arguing over $k$ (or $m$)'', simply asserting $m=1$ cannot be better. In this sense, the approach described here is strictly more credible than DID, and certainly more transparent in relaying how the results depend upon the assumptions.

\begin{figure}[h!]
\begin{center}
\caption{\label{f_NSWD_LinePlot} Line plot for NSW example. $\hat{\beta}_{P\sim D| Z,X}=0$}
\includegraphics[scale=0.18]{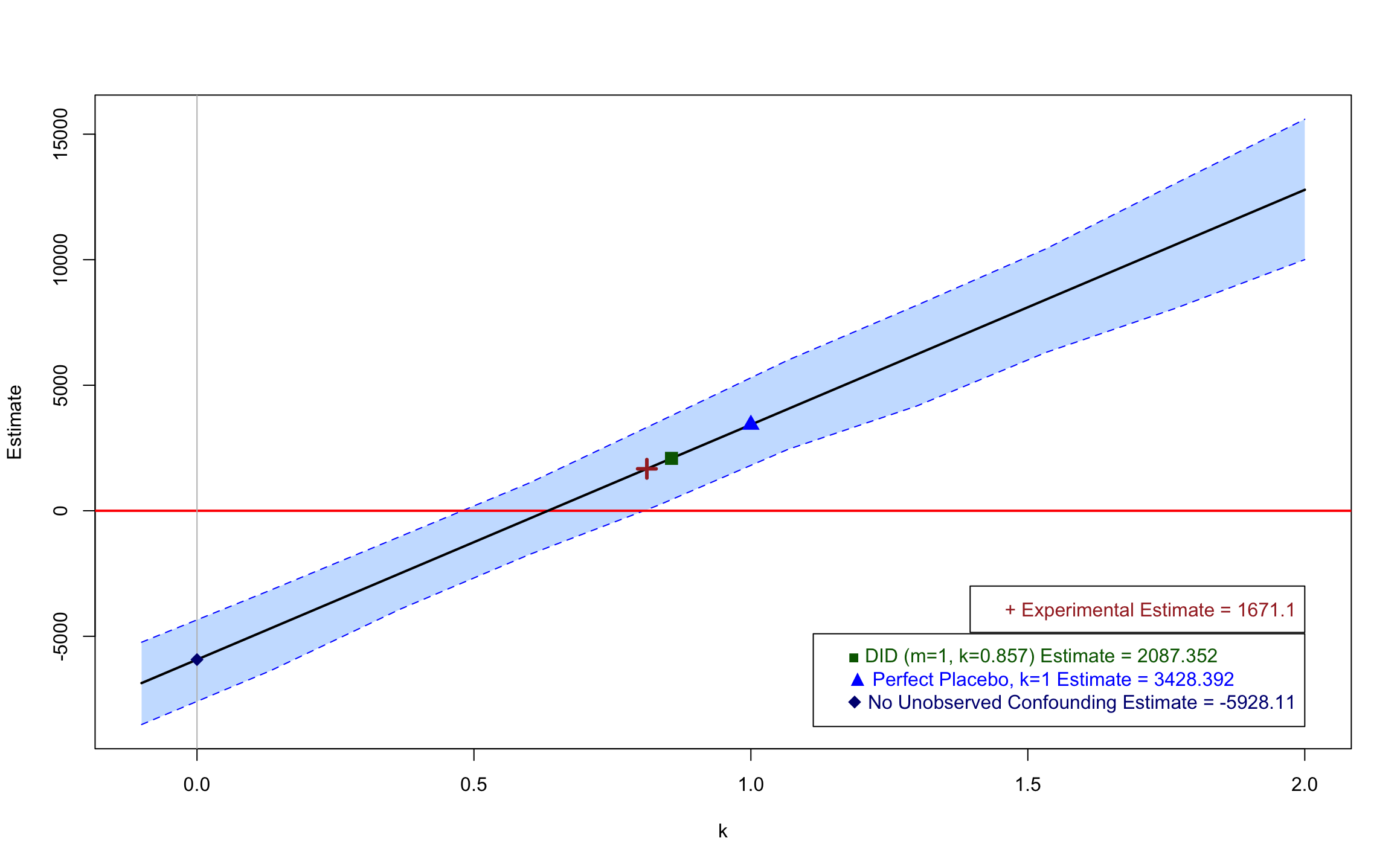}
\end{center}
\end{figure}
Figure~\ref{f_NSWD_LinePlot} shows results from a model that includes observed covariates (age, education, black, hispanic, married, nodegree). The adjusting-for-observables estimate is negative here, a well-known result when using regression methods on this dataset. Absent knowledge of the experimental benchmark, assuming only that $0.5<k<1$ places the results between -1,249 USD and +3,428 USD. Note that instead using 1974 earnings as the placebo outcome yields similar results (see Figure~\ref{f_NSWD_LinePlot74} in Appendix~\ref{code}).  In that case, $0.5<k<1$ would bound the results between -1,406 USD and +3,115 USD. The intersection of these ranges would thus place the effect between -1,249 USD and +3,115 USD. These results produce a somewhat informative range of estimates, with the estimates either being statistically insignificant towards the lower values of $k$ or positive and statistically significant once $k$ exceeds about 0.8. However, we emphasize that the value of sensitivity and partial-identification approaches lies in their ability to map weaker, more defensible assumptions to a range of estimates thereby avoiding over-confidence rooted in indefensible assumptions.  As wide as the resulting intervals are, they can be defended, and in this case they contrast sharply with the adjustment-for-observables point estimate of below -5,000 USD.  

We can also look at several point estimates of interest. First, under standard DID ($m=1$), the effect of job training estimated under this assumption is an increase of 2,087 USD--- close to the benchmark experimental estimate (1,671 USD). The $k=1$ version of DID accounts for scale changes in income, and also produces a reasonable estimate (3,428 USD). While the conventional DID result turns out to be closer, this appears to be due to two countervailing violations of parallel trends. First, because we can reason that $\beta_{P\sim D| Z, X}=0$ here, we can use the experimental benchmark to backout an estimate of the ``true'', otherwise unobservable, value of $k$ at $k=0.812$. That is, having put 1975 and 1978 earnings on the same scale, the difference in 1978 outcomes due to confounding bias is 81\% as big as the difference between the groups in 1975 (both controlling for $X$). This comports with the reasoned expectations above placing $k$ between 0.5 and 1. However, as it happens, since the scale of the outcome variable grew by 1978 ($\text{SF} = 1.167$), the ``true'' value for $m$, the ratio of the biases without adjusting for scale differences, is 0.948 ($m = k\times\text{SF} = 0.812 \times 1.167 = 0.948$). Thus, standard DID comes out close to the experimental benchmark in this example, but not out of the credibility of its assumptions.

\begin{figure}[h!]
\begin{center}
\caption{\label{f_NSWD_LinePlotu} Line plot for NSW example using employment rather than earnings as the placebo. $\beta_{P\sim D| Z,X}=0$}
\includegraphics[scale=0.18]{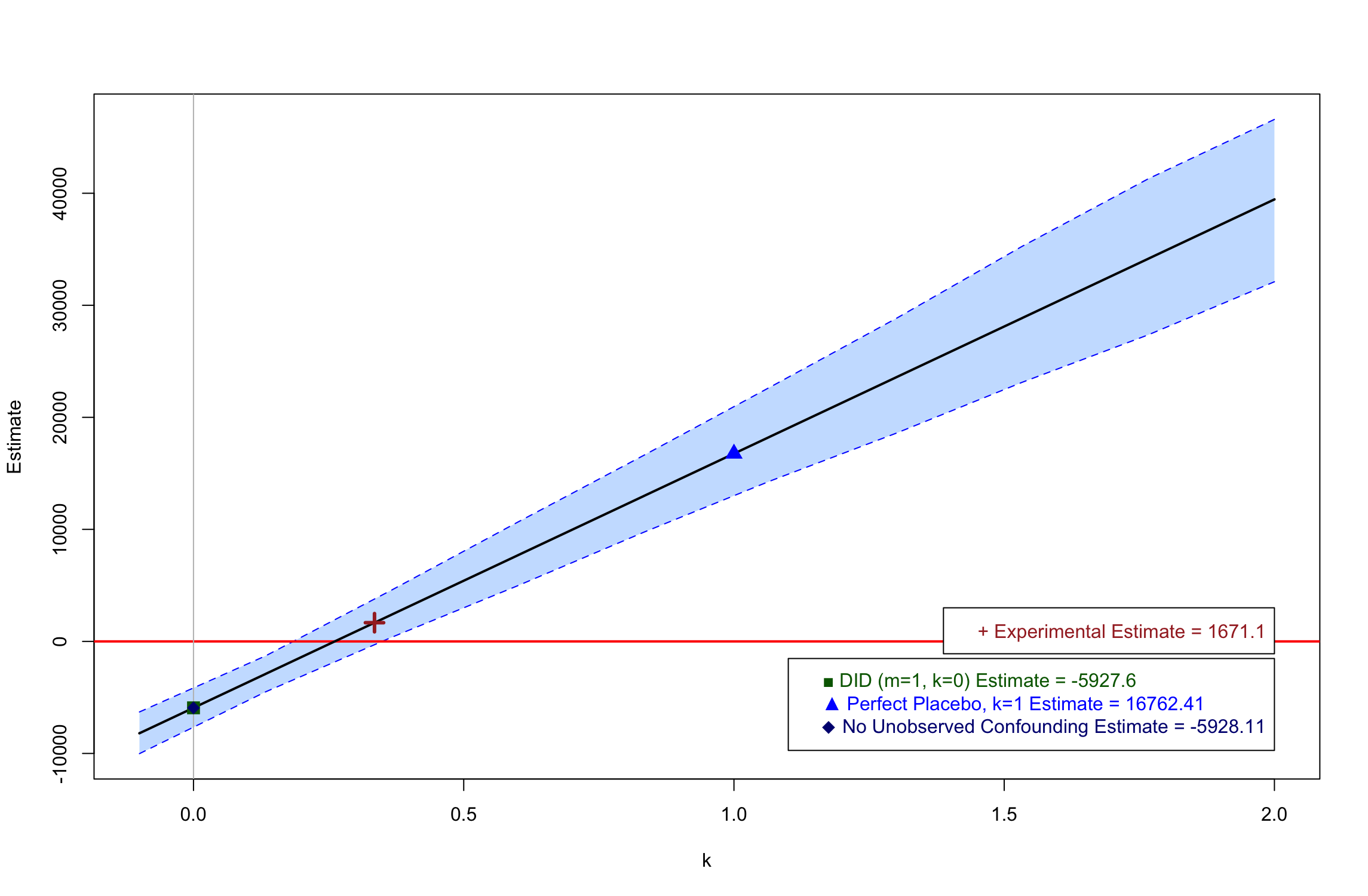}
\end{center}
\end{figure}

\paragraph{Using unemployment as the placebo outcome.}We also have data on whether or not each participant was employed in 1975 (and 1974). This is a binary variable that is simply a function of earnings, as the variable was constructed to equal one when earnings was reported to be zero. We demonstrate here how this could be used as a placebo outcome as well, despite being on a very different scale from earnings in 1978 (the scale factor exceeds 40,000). However, our reasoning about $k$ must take into account that unemployment is much coarser than income and so there must be information loss in terms of its relationship to confounders. 
In this case, we might expect that the biases have the same sign (if we code ``employment'' rather then ``unemployment''), so we expect $k>0$. Additionally, we might think 1975 employment is more confounded since selection into treatment is based on employment and earnings prior to when the program started; thus, we expect $0<k<1$. The resulting estimates are shown in Figure~\ref{f_NSWD_LinePlotu}. Note that conventional DID ($m=1$), or reasoning directly about $m$ in any way, makes little sense here given the different scales of the placebo outcome and real outcome. 
While this exercise is less informative than when we use 1975 earnings as the placebo, it still provides a way to plausibly contain the true treatment effect. Empirically, given the benchmark available in this case, we find that the ``true $k$'' would be 0.33.  This example illustrates the general consideration that investigators may have better and worse choices of placebos, depending on which can support arguments about relative confounding ($k$). In some settings we may only have mediocre choices but defensible estimates can still be revealed, given what we are and are not equipped to argue. 

\subsection{Estimating the effects of Zika virus on birth rates in Brazil}

In a second application, we investigate the effect Zika virus had on birth rates in Brazil. Zika virus can cause birth complications, which may lead either to couples delaying planned pregnancies, or to increased risk of miscarriage. We use data previously analyzed by \cite{Taddeo2022} and \cite{Tchetgen2023_single}, also employing placebo-based techniques, but requiring different assumptions. In Brazil in 2015, the state of Pernambuco saw high rates of Zika, while zero cases were reported in the state of Rio Grande do Sul. As in previous works, we regard the former state as the treated group and the later as the control group. We use municipality level overall birth rates per 1000 people in 2016 as our primary outcome of interest, while the same measure in 2014 is our placebo outcome. 

We make the reasonable assumption that the treatment has no direct causal effect on the pre-treatment (placebo) outcome.  Using data provided by \cite{Amorim2022}, we perform the analysis using the \texttt{PlaceboLM} package; see Appendix~\ref{code} for code.

\begin{figure}[ht!]
\begin{center}
\caption{\label{f_Zika_LinePlot} Line plot for Zika virus example with 2014 birth rate as placebo outcome.  $\beta_{P\sim D| Z,X}=0$}
\includegraphics[scale=0.18]{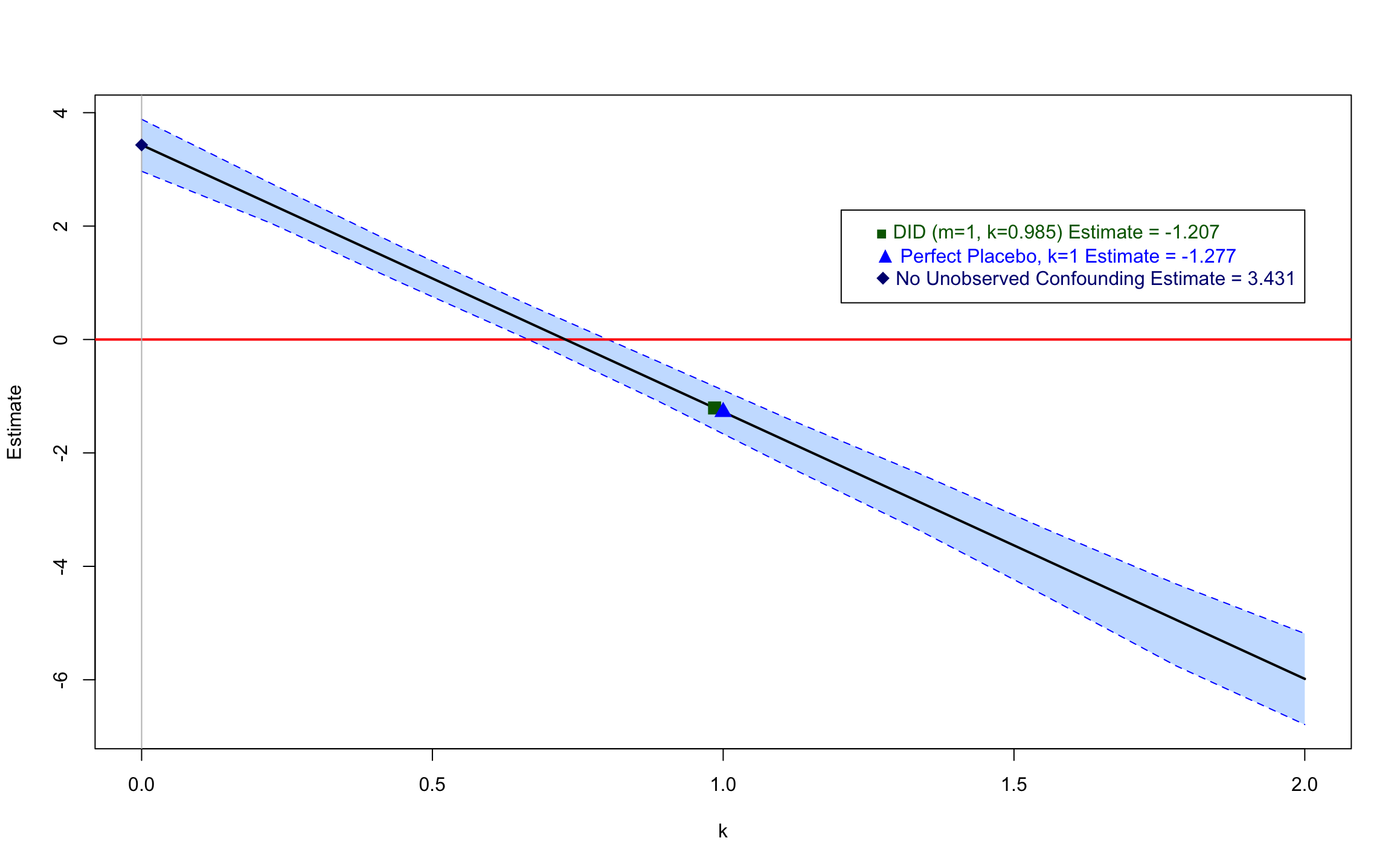}
\end{center}
\end{figure}

The naive difference in means (labeled as ``no unobserved confounding'') estimate finds that Zika is associated with 3.4 \textit{additional} births per 1000 people. By contrast, the two DID estimates ($m=1$ and $k=1$) are negative, indicating 1.2-1.3 fewer births per thousand people in Pernambuco (the Zika-treated state). 

Here again, we assume that $k>0$ as we expect any differences between the states in 2014 to have the same sign as the non-treatment difference between the states come 2018. In this case there is little reason to argue that $k<1$---unlike the NSW case, the ``selection into treatment'' was proximate in time to the post-treatment outcome. Consequently we may suspect that $k>1$ is more likely, even if we cannot rule out $k<1$ as possible.

Plotting the results over the range $0<k<2$ (Figure~\ref{f_Zika_LinePlot}) we see that if the level of confounding of 2016 birth rate and treatment is approximately 70\% of the level of confounding of 2014 birth rate and treatment or more ($k> 0.7$), then the effect estimates are negative. The estimate reaches a 2.0 births per thousand reduction when one assumes the relative level of confounding is roughly 120\% ($k=1.2$).

Our recommended interpretation would be that, in contrast to the naive comparison across states in 2016, we cannot rule out that Zika exposure led to fewer births. Further over much of the \textit{ex ante} plausible range ($k>0.7$) the data are consistent with a reduction in births. Any investigator seeking to argue that Zika did \textit{not} cause a reduction in births must then contend with the difficulty of arguing that $k<0.7$. These findings are consistent with the results in \cite{Taddeo2022} and \cite{Tchetgen2023_single}

\section{Discussion}

Efforts to employ placebos to gain causal leverage are not new.
\cite{Lipsitch2010} provide an early discussion of the potential of using placebo treatments and outcomes. \cite{Arnold2016} and \cite{Arnold2016b} discuss the potential to use placebos to address sample selection and measurement bias as well as bias in randomized experiments. \cite{Shi2020} review many established procedures and recent developments in this growing literature. 
An important related area of research is the proximal causal inference framework (see \citealp{Tchetgen2020} for an introduction). This line of work (initially) leverages ``double-negative control designs'' \citep{Shi2020} in which both a placebo outcome and a placebo treatment are required. Under certain assumptions and conditions, papers such as \cite{Miao2018,Miao2020} and \cite{Mastouri2021} show non-parametric identification results when both of these placebos are available.\footnote{In Appendix \ref{double_placebos}, we show a similar point identification result for regression when both a ``perfect'' placebo treatment and a ``perfect'' placebo outcome are available. We also relax the ``double placebo'' setting to allow for imperfect placebos.} Most critically, these assumptions include both ``perfect placebo'' type assumptions, and the use of a ``bridge function'' assumption, which is tantamount to an equiconfounding assumption on a transformation of the placebo variable. Also notable is recent work \citep{Tchetgen2023_single} in the single-placebo setting. One difference between that work and ours lies in the causal structures to which it can be applied. Specifically, that approach applies where the outcome variable and placebo are associated only due to the effect of the treatment. This rules out cases where the placebo and true outcome share a common cause confounder, as well as all other DAGs in Tables~\ref{taxonomy_table1} and ~\ref{taxonomy_table2}. It would, however, be a good fit in contexts akin to classical measurement error, such as when the placebo is a descendant of the outcome and not of any confounding variables.

While this approach applies in settings that do not resemble DID, the connection to DID is notable and relates it to a growing literature and range of ways in which investigators might consider relaxing the parallel trends assumption. Prior work has also noted the connection between placebo outcomes and DID, such as \citep{Sofer2016}, which also proposes a method to employ placebo outcomes on different scales from the true outcome. 
Scale aside, the parallel trends assumption is frequently difficult to support, however: if units may select into treatment on the basis of unobserved factors that impact the outcomes, it is difficult to then argue that no unobserved factor related to treatment assignment influences the average trend in the (non-treatment) outcome. 
Thus, additional tools that allow reasoned relaxations of the parallel trends assumptions are highly valuable. Numerous other approaches to relaxing the parallel trends assumption, or to improve on the current practice of testing for parallel trends in the pre-treatment period have been proposed (see e.g. \citealp{Manski2018,Bilinski2018,Freyaldenhoven2019,Keele2019,Ryan2019,Ye2020,Gibson2021,rambachan2023more}). 
The principal difference between any of these approaches and ours (when applied to DID settings) lies in what they ask the user to reason about.  For example, \cite{rambachan2023more} suggest that users consider violations of parallel trends to have magnitudes that are multiples of how different the treated and control group were on trends entirely prior to treatment.   \cite{Gibson2021} applies the logic of omitted variable bias as in \cite{CinelliHazlett2020} to DID, by using a first-difference regression when data are panel structured. This asks users to reason about how much of (i) the overtime change in the (non-treatment) and (ii) the overtime change in treatment activation are explained by unobserved confounding.  Our approach requires investigators to reason about how much bias is suffered in the post-treatment outcome difference, as a multiple of the bias suffered in a placebo (pre-treatment) outcome difference (i.e. $m$, or $k$). Our approach also does not require that the placebo outcome is the same variable or on the same scale, as required for pre-treatment outcomes in DID. Which of these approaches offers users the greatest leverage is a question of the specific domain and the information available. Our view is that whichever approach facilitates understanding and determining plausible bounds in a specific context is to be preferred. For example, investigators may be better equipped to argue that the confounding experienced by the pre-treatment outcome is between 0.5 and 2 times the confounding experienced by the post-treatment outcome in a context where they are not able to bound the differential trends in simple additive terms. By contrast, if the trends in the outcome absent treatment can be constrained by domain knowledge or physical limitations, then arguments about the additive difference in non-treatment trends may be more useful.
Further, we argue that any approach that shows results as a function of an unknown ``identification parameter'' (like $m$ or $k$) can be valuable and transparency-improving, simply because it reminds the reader that conventional DID estimates stand on a heroic assumption. That is, while users may argue they cannot defend any particular value of $m$ or $k$ in our setting, simply ignoring this fact by asserting that $m=1$ (as in conventional DID) is not credibility-enhancing.

The novelty of our approach lies in using the OVB framework to relax those assumptions of perfect placebos and equiconfounding. The main limitations are, first, our dependency on linearity (though see Appendix \ref{PLandNPAppendix} for partially linear and non-parametric extensions).  Second, researchers may reasonably raise the practical concern of whether the approach will ultimately produce informative estimates. As with other partial identification approaches, these tools may leave researchers unable to defend informative ranges of estimates in some or perhaps most cases. This, in our view, is a feature not a bug: When we are unable produce a sufficiently narrow or informative range of estimates, this reflects that the available data and defensible assumptions do not yet tell us as much as we would like. In this way we hope our tools aid users in producing and communicating claims about what can and cannot be safely concluded about important causal questions under imperfect identification conditions.

\singlespacing
\bibliographystyle{plainnat}
\bibliography{references.bib}

\newpage
\appendix
\setcounter{page}{1} 
\pagenumbering{arabic} 
\begin{center}
    \LARGE{\textbf{Online Appendix: }\\ 
    \Large{Causal progress with imperfect placebo treatments and outcomes}}
\end{center}

\newpage

\begin{landscape}

\addtolength{\oddsidemargin}{-.7in}%
\addtolength{\evensidemargin}{-1.2in}%
\addtolength{\textwidth}{1.5in}%

{\tiny
\begin{table}[h!]
\caption{\label{taxonomy_table1} {\bf Taxonomy of Single Placebo Graphs} }


\centering
\begin{tabular}{|c|p{2in}|p{0.75in}|c|l|p{2in}|}
\hline
 &   Graph ($\mathbf{X}$ omitted for clarity) &  Placebo Type & Short Regression(s) & Parameters & Target Coefficient Expression 
\\ \hline
[a] 
&
\multirow{2}{2in}{
\begin{tikzpicture}[baseline=(current bounding box.north)]
\node (D) at (0,0) {$D$};
\node (Y) at (4,0) {$Y$};
\node (P) at (2,1) {$P$};
\node (Z) at (2,2) {$\mathbf{Z}$};
\path (D) edge (Y);
\path (Z) edge (D);
\path (Z) edge (Y);
\path (Z) edge (P);
\end{tikzpicture}
}
& 
Placebo Outcome
&  
$\begin{aligned}
Y = 
&\beta_{Y\sim D | X} D + \\
&\mathbf{X}\beta_{Y \sim X| D} + \epsilon_s \\
P = 
&\beta_{P\sim D | X} D + \\
&\mathbf{X}\beta_{P \sim X| D} + \xi_s
\end{aligned}$
&
$\begin{aligned}
&k = \frac{\text{bias}_{\text{(YD.X)}} }{\text{bias}_{\text{(PD.X)}} \times \text{SF}} \\
&\beta_{P\sim D | X,Z} (= 0 \text{ in this graph}) \\
&\text{SF} = \frac{\sd(Y^{\perp D,X})}{\sd(P^{\perp D,X})}
\end{aligned}$
&
$\begin{aligned}
&\beta_{Y\sim D| Z, X} = \beta_{Y\sim D | X} - \\
&k  \times \left(\beta_{P\sim D | X} - \beta_{P\sim D| Z, X} \right)  \\
&\times \text{SF}
\end{aligned}$

\\ \cline{3-6}

&

&   
Placebo Treatment
&   
$\begin{aligned}
Y = 
&\beta_{Y\sim D | P,X} D + \\
&\beta_{Y\sim P| D, X} P + \\
&\mathbf{X}\beta_{Y \sim X| D,P} + \epsilon_s
\end{aligned}$
&
$\begin{aligned}
&k = \frac{\text{bias}_{\text{(YD.PX)}}}{\text{bias}_{\text{(YP.DX)}}\times \text{SF}} \\
&\beta_{Y\sim P| D,Z, X}  (= 0 \text{ in this graph})\\
&\text{SF} = \frac{\sd(P^{\perp D,X})}{\sd(D^{\perp P,X})}
\end{aligned}$
&
$\begin{aligned}
&\beta_{Y\sim D| P, Z, X} = \beta_{Y\sim D | P,X} - \\
&k \times \left(\beta_{Y\sim P | D,X} - \beta_{Y\sim P| D,Z, X}\right)  \\
&\times  \text{SF}
\end{aligned}$

\\ \cline{1-6}
[b]
&
\begin{tikzpicture}[baseline=(current bounding box.north)]
\node (D) at (0,0) {$D$};
\node (Y) at (4,0) {$Y$};
\node (P) at (2,1) {$P$};
\node (Z) at (2,2) {$\mathbf{Z}$};
\path (D) edge (Y);
\path (Z) edge (D);
\path (Z) edge (Y);
\path (Z) edge (P);
\path (D) edge (P);
\end{tikzpicture}
&   
\multicolumn{4}{l}{Placebo Outcome (see above row); $\beta_{P\sim D | X,Z} \ne 0$} \vline

\\ \cline{1-6}
[c]
&
\multirow{2}{2in}{
\begin{tikzpicture}[baseline=(current bounding box.north)]
\node (D) at (0,0) {$D$};
\node (Y) at (4,0) {$Y$};
\node (P) at (2,1) {$P$};
\node (Z) at (2,2) {$\mathbf{Z}$};
\path (D) edge (Y);
\path (Z) edge (D);
\path (Z) edge (Y);
\path (Z) edge (P);
\path (P) edge (Y);
\end{tikzpicture}
}

&   
\multicolumn{4}{l}{Placebo Treatment (see above row); $\beta_{Y\sim P| D,Z, X} \ne 0$} \vline

\\ \cline{3-6}

&

&
Observed Confounder 1
&   
$\begin{aligned}
Y = 
&\beta_{Y\sim D | P, X} D + \\
&\beta_{Y\sim P | D, X} P + \\
&\mathbf{X}\beta_{Y \sim X| D, P} + \epsilon_s \\
P = 
&\beta_{P\sim D | X} D + \\
&\mathbf{X}\beta_{P \sim X| D} + \xi_s
\end{aligned}$
&   
$\begin{aligned}
&k =
\frac
{\text{bias}_{\text{(YD.PX)}} }
{\text{bias}_{\text{(PD.X)}}\times \text{SF}} \\
&\beta_{P\sim D| Z , X} (= 0 \text{ in this graph}) \\
&\text{SF} = \frac{\sd(Y^{\perp D,P,X})}{\sd(D^{\perp P,X})}
\times \frac{\sd(D^{\perp X})}{\sd(P^{\perp D,X})}
\end{aligned}$
&
$\begin{aligned}
&\beta_{Y\sim D| P, Z, X} = \beta_{Y\sim D | P,X} - \\
&k \times (\beta_{P\sim D | X} - \beta_{P\sim D| Z , X} ) \\
&\times \text{SF} 
\end{aligned}$

\\ \cline{1-6}
[d]
&
\multirow{4}{2in}{
\begin{tikzpicture}[baseline=(current bounding box.north)]
\node (D) at (0,0) {$D$};
\node (Y) at (4,0) {$Y$};
\node (P) at (2,1) {$P$};
\node (Z) at (2,2) {$\mathbf{Z}$};
\path (D) edge (Y);
\path (Z) edge (D);
\path (Z) edge (Y);
\path (Z) edge (P);
\path (D) edge (P);
\path (P) edge (Y);
\end{tikzpicture}
}
&   
\multicolumn{4}{l}{See \cite{Zhang2022} for partial identification of indirect and direct effects} 
\vline

\\ \cline{3-6}

&

&   
\multicolumn{4}{l}{See \cite{CinelliHazlett2020} for partial ID of total effects; $R^2_{Y\sim Z|D,X}$ includes $Z \to P \to Y$} 
\vline

\\ \cline{3-6}

&

&   
\multicolumn{4}{l}{Placebo outcome (see above); $\beta_{P\sim D | X,Z}$ is part of total effect; $\text{bias}_{\text{(YD.X)}}$ includes $Z \to P \to Y$} 
\vline

\\ \cline{3-6}

&

&
Mediator
&   
$\begin{aligned}
Y = 
&\beta_{Y\sim D | X} D + \\
&\mathbf{X}\beta_{Y \sim X| D} + \epsilon_s \\
Y = 
&\beta_{Y\sim D | P,X} D + \\
&\beta_{Y\sim P| D, X} P + \\
&\mathbf{X}\beta_{Y \sim X| D,P} + \xi_s
\end{aligned}$
&
$\begin{aligned}
&k =
\frac{{\text{bias}_{\text{(YD.X)}}}}{{\text{bias}_{\text{(YP.DX)}}}\times \text{SF} } \\
&(\text{bias}_{\text{(YD.X)}} \text{ includes $Z \to P \to Y$}) \\
&\beta_{Y\sim P| D, Z, X} \text{ is part of total effect} \\
&\text{SF} = \frac{\sd(P^{\perp D,X})}{\sd(D^{\perp X})} \frac{\sd(Y^{\perp D,X})}{\sd(Y^{\perp D,P,X})}
\end{aligned}$
&
$\begin{aligned}
&\beta_{Y\sim D|  Z, X}
= \beta_{Y\sim D | X} - \\
&k \times ( \beta_{Y\sim P | D,X} - \beta_{Y\sim P| D, Z, X}  ) \\
&\times \text{SF} 
\end{aligned}$

\\ \hline
\end{tabular}
\end{table}
}
\end{landscape}


\begin{landscape}
{\tiny
\begin{table}[h!]
\caption{\label{taxonomy_table2} {\bf Taxonomy of Single Placebo Graphs - Continued}}


\centering
\begin{tabular}{|c|p{2in}|p{0.75in}|c|l|p{2in}|}
\hline
 &   Graph  ($\mathbf{X}$ omitted for clarity) &  Placebo Type & Short Regression(s) & Parameters & Target Coefficient Expression 
\\ \hline
[e] 
&
\begin{tikzpicture}[baseline=(current bounding box.north)]
\node (D) at (0,0) {$D$};
\node (Y) at (4,0) {$Y$};
\node (P) at (2,1) {$P$};
\node (Z) at (2,2) {$\mathbf{Z}$};
\path (D) edge (Y);
\path (Z) edge (D);
\path (Z) edge (Y);
\path (Z) edge (P);
\path (P) edge (D);
\end{tikzpicture}
& 
Observed Confounder 2
&  
$\begin{aligned}
Y = 
&\beta_{Y\sim D | P,X} D + \\
&\beta_{Y\sim P| D, X} P + \\
&\mathbf{X}\beta_{Y \sim X| D,P} + \epsilon_s \\
D = 
&\beta_{D\sim P|  X} P + \\
&\mathbf{X}\beta_{D \sim X| P} + \xi_s
\end{aligned}$
&
$\begin{aligned}
&k = 
\frac{{\text{bias}_{\text{(YD.PX)}}}}{{\text{bias}_{\text{(DP.X)}}}\times \text{SF}} \\
&\beta_{D\sim P| Z , X} \ne 0  \\
&\text{SF} = \frac{\sd(Y^{\perp D,P,X})}{\sd(D^{\perp P,X})}
\times \frac{\sd(P^{\perp X})}{\sd(D^{\perp P,X})}
\end{aligned}$
&
$\begin{aligned}
&\beta_{Y\sim D| P, Z, X} = \beta_{Y\sim D | P,X} - \\
&k \times (\beta_{D\sim P | X} - \beta_{D\sim P| Z , X}) \\
&\times \text{SF} 
\end{aligned}$

\\ \cline{1-6}
[f]
&
\begin{tikzpicture}[baseline=(current bounding box.north)]
\node (D) at (0,0) {$D$};
\node (Y) at (4,0) {$Y$};
\node (P) at (2,1) {$P$};
\node (Z) at (2,2) {$\mathbf{Z}$};
\path (D) edge (Y);
\path (Z) edge (D);
\path (Z) edge (Y);
\path (Z) edge (P);
\path (P) edge (D);
\path (P) edge (Y);
\end{tikzpicture}
&   
\multicolumn{4}{l}{Observed Confounder 2 (see above row)} \vline

\\ \cline{1-6}
[g] 
&
\begin{tikzpicture}[baseline=(current bounding box.north)]
\node (D) at (0,0) {$D$};
\node (Y) at (4,0) {$Y$};
\node (P) at (2,1) {$P$};
\node (Z) at (2,2) {$\mathbf{Z}$};
\path (D) edge (Y);
\path (Z) edge (D);
\path (Z) edge (Y);
\path (Z) edge (P);
\path (Y) edge (P);
\end{tikzpicture}
& 
Post-Outcome
&  
$\begin{aligned}
Y = 
&\beta_{Y\sim D | X} D + \\
&\mathbf{X}\beta_{Y \sim X| D} + \epsilon_s \\
P = 
&\beta_{P\sim D | Y, X} D + \\
&\beta_{P\sim Y | D, X} Y + \\
&\mathbf{X}\beta_{P \sim X| D,Y} + \xi_s
\end{aligned}$
&
$\begin{aligned}
&k = \frac{{\text{bias}_{\text{(YD.X)}}}}{{\text{bias}_{\text{(PY.DX)}}}\times \text{SF}}  \\
&\beta_{P\sim Y| D, Z , X} \ne 0 \\
&\text{SF} = \frac{\sd(Y^{\perp D,X})}{\sd(D^{\perp X})}
\times \frac{\sd(Y^{\perp D,X})}{\sd(P^{\perp Y,D,X})}
\end{aligned}$
&
$\begin{aligned}
&\beta_{Y\sim D| Z, X} = \beta_{Y\sim D | X} - \\
&k \times (\beta_{P\sim Y | D,X} - \beta_{P\sim Y| D, Z , X}) \\
&\times \text{SF}
\end{aligned}$

\\ \cline{1-6}
[h]
&
\begin{tikzpicture}[baseline=(current bounding box.north)]
\node (D) at (0,0) {$D$};
\node (Y) at (4,0) {$Y$};
\node (P) at (2,1) {$P$};
\node (Z) at (2,2) {$\mathbf{Z}$};
\path (D) edge (Y);
\path (Z) edge (D);
\path (Z) edge (Y);
\path (Z) edge (P);
\path (D) edge (P);
\path (Y) edge (P);
\end{tikzpicture}
&   
\multicolumn{4}{l}{Post-Outcome (see above row)} \vline
\\ \hline
\end{tabular}
\end{table}
}
\end{landscape}

\clearpage

\setlength{\topmargin}{-1in}
\setlength{\oddsidemargin}{-.5in}
\setlength{\textwidth}{6in}
\setlength{\textheight}{7in}

\section{Single Placebo Derivations}
\label{SPDs}

\paragraph{Placebo Treatment (Table \ref{taxonomy_table1}[a])}

\begin{equation} \label{e_mult_z_pt_m_est}
\begin{aligned}
&\text{bias}_{\text{(YD.PX)}}  = \beta_{Y\sim D | P,X} - \beta_{Y\sim D| P, Z, X}
= 
 R_{Y \sim Z | D,P,X}  f_{D\sim  Z_{\text{(Y.DPX)}}|P,X}   
(\sd(Y^{\perp D,P,X})/\sd(D^{\perp P,X}))
\\
&\text{bias}_{\text{(YP.DX)}}  = \beta_{Y\sim P | D,X} - \beta_{Y\sim P| D, Z, X} 
= 
 R_{Y \sim Z | D,P,X}  f_{P\sim  Z_{\text{(Y.DPX)}}|D,X}   
 (\sd(Y^{\perp D,P,X})/\sd(P^{\perp D,X})) \\
&\text{bias}_{\text{(YD.PX)}} = m \times \text{bias}_{\text{(YP.DX)}}
\implies 
\beta_{Y\sim D| P, Z, X}
= \beta_{Y\sim D | P,X} - m \times ( \beta_{Y\sim P | D,X} - \beta_{Y\sim P| D, Z, X}  ) \\
&m
= \frac{{\text{bias}_{\text{(YD.PX)}}}}{{\text{bias}_{\text{(YP.DX)}}}}
=  
\frac{R_{Y \sim Z | D,P,X}  f_{D\sim  Z_{\text{(Y.DPX)}}|P,X}}
{R_{Y \sim Z | D,P,X}  f_{P\sim  Z_{\text{(Y.DPX)}}|D,X}} 
\times \frac{\sd(P^{\perp D,X})}{\sd(D^{\perp P,X})} 
= k \times \frac{\sd(P^{\perp D,X})}{\sd(D^{\perp P,X})} \\
&\implies 
k = \frac{{\text{bias}_{\text{(YD.PX)}}}}{{\text{bias}_{\text{(YP.DX)}}}\times \frac{\sd(P^{\perp D,X})}{\sd(D^{\perp P,X})}}
\\
&\therefore 
\beta_{Y\sim D| P, Z, X}
= \beta_{Y\sim D | P,X} - k \times \left(\beta_{Y\sim P | D,X} - \beta_{Y\sim P| D,Z, X}\right) \times \frac{\sd(P^{\perp D,X})}{\sd(D^{\perp P,X})} 
\end{aligned}
\end{equation}

\paragraph{Observed Confounder 1 (Table \ref{taxonomy_table1}[c])}

\begin{equation} \label{e_p_to_y}
\begin{aligned}
&\text{bias}_{\text{(YD.PX)}}  = \beta_{Y\sim D | P,X} - \beta_{Y\sim D| P, Z, X}
= 
 R_{Y \sim Z | D,P,X}  f_{D\sim  Z_{\text{(Y.DPX)}}|P,X} 
 (\sd(Y^{\perp D,P,X}) / \sd(D^{\perp P,X})) \\
&\text{bias}_{\text{(PD.X)}}  
= \beta_{P\sim D | X} - \beta_{P\sim D| Z , X} 
= R_{P \sim Z  | D,X}  f_{D\sim  Z_{\text{(P.DX)}} |X}  
(\sd(P^{\perp D,X})/\sd(D^{\perp X})) \\
&\text{bias}_{\text{(YD.PX)}} = m \times \text{bias}_{\text{(PD.X)}} 
\implies 
\beta_{Y\sim D| P, Z, X}
= \beta_{Y\sim D | P,X} - m \times (\beta_{P\sim D | X} - \beta_{P\sim D| Z , X} ) \\
&m
= \frac{{\text{bias}_{\text{(YD.PX)}}}}{{\text{bias}_{\text{(PD.X)}} }}
=  
\frac
{R_{Y \sim Z | D,P,X}  f_{D\sim  Z_{\text{(Y.DPX)}}|P,X}}
{R_{P \sim Z  | D,X}  f_{D\sim  Z_{\text{(P.DX)}} |X}} 
\times \frac{\sd(Y^{\perp D,P,X})}{\sd(D^{\perp P,X})}
\times \frac{\sd(D^{\perp X})}{\sd(P^{\perp D,X})} 
= k
\times \frac{\sd(Y^{\perp D,P,X})}{\sd(D^{\perp P,X})}
\times \frac{\sd(D^{\perp X})}{\sd(P^{\perp D,X})}\\
&\implies 
k = \frac{{\text{bias}_{\text{(YD.PX)}}}}{{\text{bias}_{\text{(PD.X)}} } \times \frac{\sd(Y^{\perp D,P,X})}{\sd(D^{\perp P,X})}
\times \frac{\sd(D^{\perp X})}{\sd(P^{\perp D,X})}}
\\
&\therefore
\beta_{Y\sim D| P, Z, X}
= \beta_{Y\sim D | P,X} - k \times (\beta_{P\sim D | X} - \beta_{P\sim D| Z , X} ) 
\times \frac{\sd(Y^{\perp D,P,X})}{\sd(D^{\perp P,X})}
\times \frac{\sd(D^{\perp X})}{\sd(P^{\perp D,X})}
\end{aligned}
\end{equation}

\newpage

\paragraph{Mediator (Table \ref{taxonomy_table1}[d])}

\begin{equation} \label{e_mult_z_med}
\begin{aligned}
&\text{bias}_{\text{(YD.X)}}  = \beta_{Y\sim D | X} - \beta_{Y\sim D|  Z, X}
= 
 R_{Y \sim Z | D,X}  f_{D\sim  Z_{\text{(Y.DX)}}|X}   
 (\sd(Y^{\perp D,X})/\sd(D^{\perp X}))
\\
&\text{bias}_{\text{(YP.DX)}}  = \beta_{Y\sim P | D,X} - \beta_{Y\sim P| D, Z, X} 
= 
 R_{Y \sim Z | D,P,X}  f_{P\sim  Z_{\text{(Y.DPX)}}|D,X}   
 (\sd(Y^{\perp D,P,X})/\sd(P^{\perp D,X})) \\
&\text{bias}_{\text{(YD.X)}} = m \times \text{bias}_{\text{(YP.DX)}} 
\implies 
\beta_{Y\sim D|  Z, X}
= \beta_{Y\sim D | X} - m \times ( \beta_{Y\sim P | D,X} - \beta_{Y\sim P| D, Z, X}  ) \\
&m
= \frac{{\text{bias}_{\text{(YD.X)}}}}{{\text{bias}_{\text{(YP.DX)}}}}
=  
\frac{R_{Y \sim Z | D,X}  f_{D\sim  Z_{\text{(Y.DX)}}|X}}
{R_{Y \sim Z | D,P,X}  f_{P\sim  Z_{\text{(Y.DPX)}}|D,X}} 
 \frac{\sd(P^{\perp D,X})}{\sd(D^{\perp X})} 
\frac{\sd(Y^{\perp D,X})}{\sd(Y^{\perp D,P,X})} 
= k \times \frac{\sd(P^{\perp D,X})}{\sd(D^{\perp X})} \frac{\sd(Y^{\perp D,X})}{\sd(Y^{\perp D,P,X})}  \\
&\implies 
k = \frac{{\text{bias}_{\text{(YD.X)}}}}{{\text{bias}_{\text{(YP.DX)}}}\frac{\sd(P^{\perp D,X})}{\sd(D^{\perp X})} \frac{\sd(Y^{\perp D,X})}{\sd(Y^{\perp D,P,X})} }
\\
&\therefore
\beta_{Y\sim D|  Z, X}
= \beta_{Y\sim D | X} - k \times ( \beta_{Y\sim P | D,X} - \beta_{Y\sim P| D, Z, X}  ) \times \frac{\sd(P^{\perp D,X})}{\sd(D^{\perp X})} \frac{\sd(Y^{\perp D,X})}{\sd(Y^{\perp D,P,X})} 
\end{aligned}
\end{equation}

\paragraph{Observed Confounder 2 (Table \ref{taxonomy_table2}[e])}

\begin{equation} \label{e_obs_conf}
\begin{aligned}
&\text{bias}_{\text{(YD.PX)}}  = \beta_{Y\sim D | P,X} - \beta_{Y\sim D| P, Z, X}
= 
 R_{Y \sim Z| D,P,X}  f_{D\sim  Z_{\text{(Y.DPX)}}|P,X} 
 (\sd(Y^{\perp D,P,X}) / \sd(D^{\perp P,X})) \\
&\text{bias}_{\text{(DP.X)}}  
= \beta_{D\sim P | X} - \beta_{D\sim P| Z , X} 
= R_{D \sim Z  | P,X}  f_{P\sim  Z_{\text{(D.PX)}} |X}  
(\sd(D^{\perp P,X})/\sd(P^{\perp X})) \\
&\text{bias}_{\text{(YD.PX)}} = m \times \text{bias}_{\text{(DP.X)}} 
\implies 
\beta_{Y\sim D| P, Z, X}
= \beta_{Y\sim D | P,X} - m \times (\beta_{D\sim P | X} - \beta_{D\sim P| Z , X}) \\
&m
= \frac{{\text{bias}_{\text{(YD.PX)}}}}{{\text{bias}_{\text{(DP.X)}}}}
= \frac
{R_{Y \sim Z | D,P,X}  f_{D\sim  Z_{\text{(Y.DPX)}}|P,X} }
{R_{D \sim Z  | P,X}  f_{P\sim  Z_{\text{(D.PX)}} |X}} 
\times \frac{\sd(Y^{\perp D,P,X})}{\sd(D^{\perp P,X})}
\times \frac{\sd(P^{\perp X})}{\sd(D^{\perp P,X})} 
= k
\times \frac{\sd(Y^{\perp D,P,X})}{\sd(D^{\perp P,X})}
\times \frac{\sd(P^{\perp X})}{\sd(D^{\perp P,X})} \\
&\implies
k = 
\frac{{\text{bias}_{\text{(YD.PX)}}}}{{\text{bias}_{\text{(DP.X)}}}\times \frac{\sd(Y^{\perp D,P,X})}{\sd(D^{\perp P,X})} \frac{\sd(P^{\perp X})}{\sd(D^{\perp P,X})}}
\\
&\therefore
\beta_{Y\sim D| P, Z, X}
= \beta_{Y\sim D | P,X} - k \times (\beta_{D\sim P | X} - \beta_{D\sim P| Z , X})
\times \frac{\sd(Y^{\perp D,P,X})}{\sd(D^{\perp P,X})}
\times \frac{\sd(P^{\perp X})}{\sd(D^{\perp P,X})}
\end{aligned}
\end{equation}

\paragraph{Post-outcome (Table \ref{taxonomy_table2}[g])}

\begin{equation} \label{e_post_outcome}
\begin{aligned}
&\text{bias}_{\text{(YD.X)}}  = \beta_{Y\sim D | X} - \beta_{Y\sim D| Z, X}
= R_{Y \sim Z | D,X}  f_{D\sim  Z_{\text{(Y.DX)}}|X} 
 (\sd(Y^{\perp D,X}) / \sd(D^{\perp X})) \\
 &\text{bias}_{\text{(PY.DX)}}  
= \beta_{P\sim Y | D,X} - \beta_{P\sim Y| D, Z , X} 
= R_{P \sim Z  | Y,D,X}  f_{Y\sim  Z_{\text{(P.YDX)}} |D,X}  
(\sd(P^{\perp Y,D,X})/\sd(Y^{\perp D,X})) \\
&\text{bias}_{\text{(YD.X)}} = m \times \text{bias}_{\text{(PY.DX)}} 
\implies 
\beta_{Y\sim D| Z, X}
= \beta_{Y\sim D | X} - m \times (\beta_{P\sim Y | D,X} - \beta_{P\sim Y| D, Z , X}) \\
&m
= \frac{{\text{bias}_{\text{(YD.X)}}}}{{\text{bias}_{\text{(PY.DX)}}}}
= \frac
{R_{Y \sim Z | D,X}  f_{D\sim  Z_{\text{(Y.DX)}}|X}  }
{R_{P \sim Z | Y,D,X}  f_{Y\sim  Z_{\text{(P.YDX)}} |D,X}  } 
\times \frac{\sd(Y^{\perp D,P,X})}{\sd(D^{\perp P,X})}
\times \frac{\sd(P^{\perp X})}{\sd(D^{\perp P,X})} 
= k 
\times \frac{\sd(Y^{\perp D,X})}{\sd(D^{\perp X})}
\times \frac{\sd(Y^{\perp D,X})}{\sd(P^{\perp Y,D,X})} \\
&\implies 
k = 
\frac{{\text{bias}_{\text{(YD.X)}}}}{{\text{bias}_{\text{(PY.DX)}}}\times \frac{\sd(Y^{\perp D,X})}{\sd(D^{\perp X})}
\times \frac{\sd(Y^{\perp D,X})}{\sd(P^{\perp Y,D,X})}}
\\
&\therefore
\beta_{Y\sim D| Z, X}
= \beta_{Y\sim D | X} - k \times (\beta_{P\sim Y | D,X} - \beta_{P\sim Y| D, Z , X}) 
\times \frac{\sd(Y^{\perp D,X})}{\sd(D^{\perp X})}
\times \frac{\sd(Y^{\perp D,X})}{\sd(P^{\perp Y,D,X})}
\end{aligned}
\end{equation}

\section{Double placebos}
\label{double_placebos}

Consider a setting like Figure \ref{f_double_placebos} - that is, we observe both a placebo outcome ($N$) and a placebo treatment ($P$). Suppose we are interested in $\beta_{Y\sim D| Z, X}$ from the ``long'' regression in Equation \ref{e_long_reg_outcome_y}. However, $Z$ is unobserved and so we must only consider estimating ``short'' regression in Equation \ref{e_short_reg_outcome_y}. 
We might consider the following two sets of short and long regressions. Again, $\mathbf{Z}$ is a set of unobserved confounders. We can only consider estimating the short regressions (i.e., Equations \ref{e_double_short_y} and \ref{e_double_short_n}). 
$Z_{\text{(Y.DPX)}} = \mathbf{Z}\beta_{Y \sim Z|D,P,X}$ is the linear combination that de-confounds the $Y,D$ and the $Y,P$ relationship. 
$Z_{\text{(N.DPX)}} = \mathbf{Z}\beta_{N \sim Z|D,P,X}$ is the linear combination that de-confounds the $N,D$ and the $N,P$ relationship. We choose $\mathbf{Z}$ to be a rich enough set of variables to allow for this interpretation.

\begin{subequations} 
\begin{align}
Y &= \beta_{Y\sim D| P, Z, X} D + \beta_{Y\sim P| D, Z, X} P + \mathbf{X}\beta_{Y \sim X| D,P,Z} +  \mathbf{Z} \beta_{Y \sim Z|D,P,X}   + \epsilon_{y,l} 
\label{e_double_long_y}\\
Y &= \beta_{Y\sim D | P,X} D + \beta_{Y\sim P| D, X} P + \mathbf{X}\beta_{Y \sim X| D,P} + \epsilon_{y,s} 
\label{e_double_short_y}\\
N &= \beta_{N\sim D| P, Z, X} D + \beta_{N\sim P| D, Z, X} P + \mathbf{X}\beta_{N \sim X| D,P,Z} + \mathbf{Z} \beta_{N\sim Z|D,P,X}  + \epsilon_{n,l} 
\label{e_double_long_n}\\
N &= \beta_{N\sim D | P,X} D + \beta_{N\sim P| D, X} P + \mathbf{X}\beta_{N \sim X| D,P} + \epsilon_{n,s} 
\label{e_double_short_n}
\end{align}
\end{subequations}

\begin{figure}[h!]
\begin{center}
\caption{\label{f_double_placebos} A single causal graph with both a placebo outcome, $N$, and with a placebo treatment, $P$. $D$ is treatment, $Y$ is outcome, and $\mathbf{Z}$ contains unobserved confounders. $\mathbf{X}$ contains observed covariates. }

\begin{tikzpicture}[baseline=(current bounding box.north)]
\node (0) at (3,5) {(a) Perfect Double Placebo};
\node (D) at (0,0) {$D$};
\node (Y) at (6,0) {$Y$};
\node (P) at (2,1) {$P$};
\node (N) at (4,1) {$N$};
\node (Z) at (3,3) {$\mathbf{Z}$};
\path (D) edge (Y);
\path (Z) edge (D);
\path (Z) edge (Y);
\path (Z) edge (P);
\path (Z) edge (N);
\path (P) edge (D);
\path (N) edge (Y);
\node[gray] (X) at (3,4) {$\mathbf{X}$};
\path[gray] (X) edge [out=240,in=90] (P);
\path[gray] (X) edge [out=300,in=90] (N);
\path[gray] (X) edge [out=320,in=90] (Y);
\path[gray] (X) edge [out=220,in=90] (D);
\path[gray,bidirected] (X) edge (Z);
\end{tikzpicture}\hspace{1cm}
\begin{tikzpicture}[baseline=(current bounding box.north)]
\node (0) at (3,5) {(a) Imperfect Double Placebo};
\node (D) at (0,0) {$D$};
\node (Y) at (6,0) {$Y$};
\node (P) at (2,1) {$P$};
\node (N) at (4,1) {$N$};
\node (Z) at (3,3) {$\mathbf{Z}$};
\path (D) edge (Y);
\path (Z) edge (D);
\path (Z) edge (Y);
\path (Z) edge (P);
\path (Z) edge (N);
\path (P) edge (D);
\path (N) edge (Y);
\path (P) edge [bend right=7] (Y);
\path (D) edge [bend right=7] (N);
\path (P) edge (N);
\node[gray] (X) at (3,4) {$\mathbf{X}$};
\path[gray] (X) edge [out=240,in=90] (P);
\path[gray] (X) edge [out=300,in=90] (N);
\path[gray] (X) edge [out=320,in=90] (Y);
\path[gray] (X) edge [out=220,in=90] (D);
\path[gray,bidirected] (X) edge (Z);
\end{tikzpicture}
\end{center}
\end{figure}

Consider the following OVB expressions, where we use the relevant linear combination for each relationship.

\begin{subequations} 
\begin{align}
\beta_{Y\sim D | P,X} - \beta_{Y\sim D| P,Z_{\text{(Y.DPX)}}, X}  
=  \beta_{Y \sim Z_{\text{(Y.DPX)}}|D,P,X}\beta_{Z_{\text{(Y.DPX)}}\sim D|P,X} 
\\
\beta_{Y\sim P | D,X} - \beta_{Y\sim P| D,Z_{\text{(Y.DPX)}}, X}  
=  \beta_{Y \sim Z_{\text{(Y.DPX)}}|D,P,X}\beta_{Z_{\text{(Y.DPX)}}\sim P|D,X}
\\
\beta_{N\sim D | P,X} - \beta_{N\sim D| P,Z_{\text{(N.DPX)}}, X}  
=  \beta_{N \sim Z_{\text{(N.DPX)}}|D,P,X}\beta_{Z_{\text{(N.DPX)}}\sim D|P,X} 
\\
\beta_{N\sim P | D,X} - \beta_{N\sim P| D,Z_{\text{(N.DPX)}}, X}  
=  \beta_{N \sim Z_{\text{(N.DPX)}}|D,P,X}\beta_{Z_{\text{(N.DPX)}}\sim P|D,X} 
\end{align}
\end{subequations}

We can re-write the middle two of these as follows. 

\begin{subequations} 
\begin{align}
\beta_{Y \sim Z_{\text{(Y.DPX)}}|D,P,X} &= \frac{\beta_{Y\sim P | D,X} - \beta_{Y\sim P| D,Z_{\text{(Y.DPX)}}, X}}{\beta_{Z_{\text{(Y.DPX)}}\sim P|D,X}} \\
\beta_{Z_{\text{(N.DPX)}}\sim D|P,X} &= \frac{\beta_{N\sim D | P,X} - \beta_{N\sim D| P,Z_{\text{(N.DPX)}}, X}}{\beta_{N \sim Z_{\text{(N.DPX)}}|D,P,X}} 
\end{align}
\end{subequations}

Now we note that there are $m_A$ and $m_B$ that satisfy the following. 

\begin{subequations} 
\begin{align}
\beta_{Z_{\text{(Y.DPX)}}\sim D|P,X} &= m_{\text{A}} \times \beta_{Z_{\text{(N.DPX)}}\sim D|P,X} \\
\beta_{Z_{\text{(Y.DPX)}}\sim P|D,X}  &= m_{\text{B}} \times \beta_{Z_{\text{(N.DPX)}}\sim P|D,X} 
\end{align}
\end{subequations}

We can derive the following expression for $\beta_{Y\sim D| P,Z_{\text{(Y.DPX)}}, X}$. 
\begin{equation} 
\begin{aligned}
\beta_{Y\sim D | P,X} - \beta_{Y\sim D| P,Z_{\text{(Y.DPX)}}, X}  
&= \beta_{Y \sim Z_{\text{(Y.DPX)}}|D,P,X}\beta_{Z_{\text{(Y.DPX)}}\sim D|P,X} \\
&=  m_{\text{A}} \times \beta_{Y \sim Z_{\text{(Y.DPX)}}|D,P,X}  \beta_{Z_{\text{(N.DPX)}}\sim D|P,X}\\
&= 
  m_{\text{A}} \times 
 \left[\frac{\beta_{Y\sim P | D,X} - \beta_{Y\sim P| D,Z_{\text{(Y.DPX)}}, X}}{\beta_{Z_{\text{(Y.DPX)}}\sim P|D,X}}\right] 
\left[\frac{\beta_{N\sim D | P,X} - \beta_{N\sim D| P,Z_{\text{(N.DPX)}}, X}}{\beta_{N \sim Z_{\text{(N.DPX)}}|D,P,X}} \right] \\
&=  \frac{ m_{\text{A}}}{ m_{\text{B}}} \times 
\frac{
(\beta_{Y\sim P | D,X} - \beta_{Y\sim P| D,Z_{\text{(Y.DPX)}}, X})
(\beta_{N\sim D | P,X} - \beta_{N\sim D| P,Z_{\text{(N.DPX)}}, X})
}
{[\beta_{N\sim P | D,X} - \beta_{N\sim P| D,Z_{\text{(N.DPX)}}, X}]}
\\
\therefore 
\beta_{Y\sim D| P,Z_{\text{(Y.DPX)}}, X} 
&= \beta_{Y\sim D | P,X} - 
\frac{ m_{\text{A}}}{ m_{\text{B}}} \times 
\frac{
(\beta_{Y\sim P | D,X} - \beta_{Y\sim P| D,Z_{\text{(Y.DPX)}}, X})
(\beta_{N\sim D | P,X} - \beta_{N\sim D| P,Z_{\text{(N.DPX)}}, X})
}
{(\beta_{N\sim P | D,X} - \beta_{N\sim P| D,Z_{\text{(N.DPX)}}, X})}
\end{aligned}
\end{equation}
\begin{equation} 
\begin{aligned}
\text{Now, }
\frac{ m_{\text{A}}}{ m_{\text{B}}}
&=
\frac{\beta_{Z_{\text{(Y.DPX)}}\sim D|P,X}}
{\beta_{Z_{\text{(N.DPX)}}\sim D|P,X}}   
\frac{\beta_{Z_{\text{(Y.DPX)}}\sim P|D,X}}
{\beta_{Z_{\text{(N.DPX)}}\sim P|D,X} }   
\\
&=
\frac{
R_{Z_{\text{(Y.DPX)}}\sim D|P,X}
R_{Z_{\text{(N.DPX)}}\sim P|D,X}
}{
R_{Z_{\text{(N.DPX)}}\sim D|P,X}
R_{Z_{\text{(Y.DPX)}}\sim P|D,X}
} 
\times 
\frac{
\sd(Z_{\text{(Y.DPX)}}^{\perp P,X})}
{\sd(Z_{\text{(N.DPX)}}^{\perp P,X})}
\frac{
\sd(Z_{\text{(N.DPX)}}^{\perp D,X})}
{\sd(Z_{\text{(Y.DPX)}}^{\perp D,X})} \\
&=
\frac{
R_{Z_{\text{(Y.DPX)}}\sim D|P,X}
R_{Z_{\text{(N.DPX)}}\sim P|D,X}
}{
R_{Z_{\text{(N.DPX)}}\sim D|P,X}
R_{Z_{\text{(Y.DPX)}}\sim P|D,X}
} 
\times 
\frac{\sd(Z_{\text{(Y.DPX)}}^{\perp P,X})}
{\sd(Z_{\text{(Y.DPX)}}^{\perp D,P,X})}
\frac{\sd(Z_{\text{(N.DPX)}}^{\perp D,P,X})}
{\sd(Z_{\text{(N.DPX)}}^{\perp P,X})}
\frac{\sd(Z_{\text{(N.DPX)}}^{\perp D,X})}
{\sd(Z_{\text{(N.DPX)}}^{\perp D,P,X})}
\frac{\sd(Z_{\text{(Y.DPX)}}^{\perp D,P,X})}
{\sd(Z_{\text{(Y.DPX)}}^{\perp D,X})} \\
&=
\frac{
R_{Z_{\text{(Y.DPX)}}\sim D|P,X}
R_{Z_{\text{(N.DPX)}}\sim P|D,X}
}{
R_{Z_{\text{(N.DPX)}}\sim D|P,X}
R_{Z_{\text{(Y.DPX)}}\sim P|D,X}
} 
\times 
\frac{
\sqrt{1-R^2_{Z_{\text{(N.DPX)}} \sim D|P,X}}
\sqrt{1-R^2_{Z_{\text{(Y.DPX)}} \sim P|D,X}}
}{
\sqrt{1-R^2_{Z_{\text{(Y.DPX)}} \sim D|P,X}}
\sqrt{1-R^2_{Z_{\text{(N.DPX)}} \sim P|D,X}}
}\\
&=
\frac{
f_{Z_{\text{(Y.DPX)}}\sim D|P,X}
f_{Z_{\text{(N.DPX)}}\sim P|D,X}
}{
f_{Z_{\text{(N.DPX)}}\sim D|P,X}
f_{Z_{\text{(Y.DPX)}}\sim P|D,X}
} \\
\therefore
\frac{ m_{\text{A}}}{ m_{\text{B}}} &=
\frac{R_{Y \sim Z_{\text{(Y.DPX)}}|D,P,X}f_{Z_{\text{(Y.DPX)}}\sim D|P,X}}
{R_{Y \sim Z_{\text{(Y.DPX)}}|D,P,X}f_{Z_{\text{(Y.DPX)}}\sim P|D,X}} 
\times 
\frac{R_{N \sim Z_{\text{(N.DPX)}}|D,P,X}f_{Z_{\text{(N.DPX)}}\sim P|D,X}}
{R_{N \sim Z_{\text{(N.DPX)}}|D,P,X}f_{Z_{\text{(N.DPX)}}\sim D|P,X}} 
\overset{\Delta}{=}
k_{\text{(YD / YP)}} \times k_{\text{(NP / ND)}}
\end{aligned}
\end{equation}

Thus, we arrive at a final expression for $\beta_{Y\sim D| P,Z_{\text{(Y.DPX)}}, X} $:

\begin{equation} 
\begin{aligned} 
&\beta_{Y\sim D| P,Z_{\text{(Y.DPX)}}, X} \\
&= \beta_{Y\sim D | P,X} - 
k_{\text{(YD / YP)}} \times k_{\text{(NP / ND)}}\times 
\frac{
(\beta_{Y\sim P | D,X} - \beta_{Y\sim P| D,Z_{\text{(Y.DPX)}}, X})
(\beta_{N\sim D | P,X} - \beta_{N\sim D| P,Z_{\text{(N.DPX)}}, X})
}
{(\beta_{N\sim P | D,X} - \beta_{N\sim P| D,Z_{\text{(N.DPX)}}, X})} 
\end{aligned}
\end{equation}

This is a fairly complicated expression that involves five partial identification or sensitivity parameters. We suspect this may be unwieldy in many practical cases. However, it may be useful when certain assumptions are defensible. But first a few notes on what this expression contains. 

\begin{itemize}
    \item $k_{\text{(YD / YP)}}$ captures the scaled ratio of the level of confounding of the $Y,D$ relationship (conditional on $P$) to the level of confounding of the $Y,P$ relationship (conditional on $D$), after re-scaling one of these biases, or the ratio of bias factors. 
    \item $k_{\text{(NP / ND)}}$ captures the scaled ratio of the level of confounding of the $N,P$ relationship (conditional on $D$) to the level of confounding of the $N,D$ relationship (conditional on $P$), after re-scaling one of these biases, or the ratio of bias factors. 
    \item $\beta_{Y\sim P| D,Z_{\text{(Y.DPX)}}, X}$ measures the causal of $P$ on $Y$ (conditional on $D$). This could further be broken down into component edges with an additional application of OVB. 
    \item $\beta_{N\sim D| P,Z_{\text{(N.DPX)}}, X}$ measures the causal of $D$ on $N$ (conditional on $P$).
    \item $\beta_{N\sim P| D,Z_{\text{(N.DPX)}}, X}$ measures the causal of $P$ on $N$ (conditional on $D$).
    \item We could also use $k_{\text{(YD / ND)}} \times k_{\text{(NP / YP)}} =
\frac{R_{Y \sim Z_{\text{(Y.DPX)}}|D,P,X}f_{Z_{\text{(Y.DPX)}}\sim D|P,X}}
{R_{N \sim Z_{\text{(N.DPX)}}|D,P,X}f_{Z_{\text{(N.DPX)}}\sim D|P,X}} 
\times 
\frac{R_{N \sim Z_{\text{(N.DPX)}}|D,P,X}f_{Z_{\text{(N.DPX)}}\sim P|D,X}}
{R_{Y \sim Z_{\text{(Y.DPX)}}|D,P,X}f_{Z_{\text{(Y.DPX)}}\sim P|D,X}} $ and have similar interpretation. 
\end{itemize}

This should have a similar feel to what we saw in the main text. In certain settings, this expression simplifies and may be more practically useful. 

\begin{enumerate}
    \item If $Z_{\text{(Y.DPX)}} = Z_{\text{(N.DPX)}}$ or both $R_{Z_{\text{(Y.DPX)}}\sim D|P,X} = R_{Z_{\text{(N.DPX)}}\sim D|P,X}$ and $R_{Z_{\text{(Y.DPX)}}\sim P|D,X} = R_{Z_{\text{(N.DPX)}}\sim P|D,X}$, then $k_{\text{(YD / YP)}} \times k_{\text{(NP / ND)}} = 1$. A simple case of this is when there is a single unobserved confounder $Z$. When we believe this may be the case, we actually do not need to reason about the level of relative confounding at all. 
    \item When we have perfect placebos (i.e., when some or all of $\beta_{Y\sim P| D,Z_{\text{(Y.DPX)}}, X}$, $\beta_{N\sim D| P,Z_{\text{(N.DPX)}}, X}$, $\beta_{N\sim P| D,Z_{\text{(N.DPX)}}, X}$ equal zero), the expression also simplifies. 
    \item When both of the above hold (e.g., we have a single unobserved $Z$ and perfect placebos), we can point identify $\beta_{Y\sim D| P,Z, X}$ without reasoning about the relative level of confounding. 
\end{enumerate}

\section{Relative bias $m$ as differences in trends ($w$) in DID}\label{app_m_to_w}

Any assumption for $m$ can be turned into an assumption about how the trends in control potential outcomes compare. An assumption about a difference in confounding might initially be stated as, for example, 
\begin{equation} \label{e_did_trasform_1}
\begin{aligned}
\underbrace{\mathbb{E}[Y_0|G=1] - \mathbb{E}[Y_0|G=0]}_{\text{Bias}_Y} &=  m \times (\underbrace{\underbrace{\mathbb{E}[P|G=1] - \mathbb{E}[P|G=0]}_{\text{DIM}_P} - \beta_{P\sim G|Z}}_{\text{Bias}_P} ) \\
\iff 
\mathbb{E}[Y_0|G=1]
&= \mathbb{E}[Y_0|G=0] + m \times (\mathbb{E}[P|G=1] - \mathbb{E}[P|G=0] - \beta_{P\sim G|Z})
\end{aligned}
\end{equation}

An assumption about a difference in trend would be something like Equation \ref{e_did_trasform_2}. 

\begin{equation} \label{e_did_trasform_2}
\begin{aligned}
\underbrace{\mathbb{E}[Y_0|G=1] - \mathbb{E}[P|G=1] }_{\text{Trend}_{G=1}} &=  w \times (\underbrace{\mathbb{E}[Y_0|G=0]- \mathbb{E}[P|G=0]}_{\text{Trend}_{G=0}} ) \\
\iff 
\mathbb{E}[Y_0|G=1]
&= \mathbb{E}[P|G=1] + w \times (\mathbb{E}[Y_0|G=0]- \mathbb{E}[P|G=0])
\end{aligned}
\end{equation}

For this exposition, we assume $\beta_{P\sim G|Z}=0$. Both Equations \ref{e_did_trasform_1} and \ref{e_did_trasform_2} provide an expression for $\mathbb{E}[Y_0|G=1]$, which, in the DID setting, is unobserved while all the other components of the expressions are estimable from the data. We can set these two expressions for $\mathbb{E}[Y_0|G=1]$ equal to each other and then solve for $w$ to get the trend assumption that corresponds to our assumption on $m$. 

\begin{equation} \label{e_did_trasform_3}
\begin{aligned}
\mathbb{E}[Y_0|G=0] + m \times (\mathbb{E}[P|G=1] - \mathbb{E}[P|G=0])
= \mathbb{E}[P|G=1] + w \times (\mathbb{E}[Y_0|G=0]- \mathbb{E}[P|G=0]) 
\end{aligned}
\end{equation}

\begin{equation} \label{e_did_trasform_4}
\begin{aligned}
\iff 
w 
&=
\frac{
\mathbb{E}[Y_0|G=0] + m \times (\mathbb{E}[P|G=1] - \mathbb{E}[P|G=0])
- \mathbb{E}[P|G=1]
}{
\mathbb{E}[Y_0|G=0]- \mathbb{E}[P|G=0]
} \\
\iff 
m
&=
\frac{
\mathbb{E}[P|G=1] + w \times (\mathbb{E}[Y_0|G=0]- \mathbb{E}[P|G=0]) - \mathbb{E}[Y_0|G=0]
}{
\mathbb{E}[P|G=1] - \mathbb{E}[P|G=0] 
}
\end{aligned}
\end{equation}

We could similarly write $m$ in terms of $w$, if we preferred to make an assumption on $w$, the difference in the trend, and wanted to know what this implies about the unequal confounding. Additionally, we know that $m = k\times \frac{\sd(Y^{\perp G})}{\sd(N^{\perp G})}$ so that we can make an assumption on $k$ and see what this implies about $w$, the trend, or make a trend assumption on $w$ and see what this implies about $k$.

\section{Partially linear and non-parametric framework}
\label{PLandNPAppendix}


If investigators are interested in partially linear of fully non-parametric approaches to estimating treatment effects, we can use the omitted variables frameworks from \cite{Chernozhukov2022} to arrive at partial identification frameworks similar to those presented in the main text for settings in which placebo outcomes are available. 

\newpage
\subsection{Partially linear framework}


An investigator may be interested in estimating a partially linear model. $Y$ is the outcome; $P$ is a placebo outcome; $D$ is the treatment; $\mathbf{X}$ is a vector of observed covariates; $\mathbf{Z}$ is a vector of unobserved covariates. We consider the following short and long partially linear models. 
\begin{subequations} 
\begin{align}
Y &= \theta_{l,Y} D + f_l(\mathbf{X},\mathbf{Z}) + \epsilon_l \label{e_part_lin_y_long} \\
Y &= \theta_{s,Y} D + f_s(\mathbf{X}) + \epsilon_s
\label{e_part_lin_y_short}
\end{align}
\end{subequations}
\begin{subequations} 
\begin{align}
P &= \theta_{l,N} D + g_l(\mathbf{X},\mathbf{Z}) + \xi_l \label{e_part_lin_n_long} \\
P &= \theta_{s,N} D + g_s(\mathbf{X}) + \xi_s
\label{e_part_lin_n_short}
\end{align}
\end{subequations}

There is some $m$ such that $\text{bias}_{\text{(YD.X)}} = m\times \text{bias}_{\text{(PD.X)}}$, where $\text{bias}_{\text{(YD.X)}} \overset{\Delta}{=} \theta_{s,Y} - \theta_{l,Y}$ and $\text{bias}_{\text{(PD.X)}} \overset{\Delta}{=} \theta_{s,P} - \theta_{l,P}$. This implies that Equation \ref{e_part_lin_m} holds. 
Using Equation \ref{e_part_lin_m} for partial identification may run into the issue that $m \overset{\Delta}{=} \frac{\text{bias}_{\text{(YD.X)}}}{\text{bias}_{\text{(PD.X)}}}$ may be difficult to interpret when $Y$ and $P$ have differing scales. 

\begin{equation} \label{e_part_lin_m}
\begin{aligned} 
\theta_{l,Y}
&= 
\theta_{s,Y} - m\times (\theta_{s,P} - \theta_{l,P})
\end{aligned}
\end{equation}

\cite{Chernozhukov2022} show that omitted variable bias in partially linear setting can be written in the following way. 

\begin{subequations} 
\begin{align}
|\theta_{s,Y} - \theta_{l,Y}|^2
= |{\text{bias}_{\text{(YD.X)}}}|^2 
&= \rho_Y^2 \times S_Y^2 \times \eta^2_{Y\sim Z|D,X}  \times \frac{\eta^2_{D\sim Z|X}}{1-\eta^2_{D\sim Z|X}}  
\\
|\theta_{s,P} - \theta_{l,P}|^2
= |{\text{bias}_{\text{(PD.X)}}}|^2 
&= \rho_P^2 \times S_P^2 \times \eta^2_{P\sim Z|D,X}  \times \frac{\eta^2_{D\sim Z|X}}{1-\eta^2_{D\sim Z|X}} 
\end{align}
\end{subequations}

$\text{BF}_{\text{(YD.X)}} \overset{\Delta}{=} \rho_Y^2\eta^2_{Y\sim Z|D,X}\frac{\eta^2_{D\sim Z|X}}{1-\eta^2_{D\sim Z|X}}$ and  $\text{BF}_{\text{(PD.X)}} \overset{\Delta}{=} \rho_P^2\eta^2_{P\sim Z|D,X}\frac{\eta^2_{D\sim Z|X}}{1-\eta^2_{D\sim Z|X}}$ are ``bias factors'' that are scaled versions of the unobserved confounding from $\mathbf{Z}$. $S_Y^2$ and $S_P^2$ are identified and therefore estimable from the observed data. 
$\eta^2_{Y\sim Z|D,X}$ and $\eta^2_{D\sim Z|X}$ are Pearson's correlation ratios (or the non-parametric $R^2$s).\footnote{$\eta^2_{D\sim Z|X} = \frac{\text{Var}(\mathbb{E}[D|Z,X]) - \text{Var}(\mathbb{E}[D|X])}{\text{Var}(D) - \text{Var}(\mathbb{E}[D|X])} = \frac{\eta^2_{D\sim Z,X} - \eta^2_{D\sim X}}{ 1- \eta^2_{D\sim X}}$. $\eta^2_{Y\sim Z|D,X}$ can be similarly interpreted.} $\eta^2_{Y\sim Z|D,X}$ is the proportion of residual variation in $Y$ explained by $Z$. $\eta^2_{D\sim Z|X}$ is the proportion of residual variation in $D$ explained by $Z$. 
$\rho_Y^2$ and $\rho_P^2$ are referred to as ``degree of adversity'' in \cite{Chernozhukov2022} and reduce the actual level of bias from the maximum possible given the $\eta^2$ parameters. 
We can now see that we can re-express $m$ as in Equation \ref{sss_partlin_m_reex}.

\begin{equation} \label{sss_partlin_m_reex}
\begin{aligned} 
|m|^2 
= \frac{|\text{bias}_{\text{(YD.X)}}|^2}{|\text{bias}_{\text{(PD.X)}}|^2} 
&= \frac{
\rho_Y^2 \times S_Y^2 \times \eta^2_{Y\sim Z|D,X}  \times \frac{\eta^2_{D\sim Z|X}}{1-\eta^2_{D\sim Z|X}}
}{
\rho_P^2 \times S_P^2 \times \eta^2_{P\sim Z|D,X}  \times \frac{\eta^2_{D\sim Z|X}}{1-\eta^2_{D\sim Z|X}} 
} \\
&= 
\frac{\rho_Y^2\times\eta^2_{Y\sim Z|D,X}  \times \frac{\eta^2_{D\sim Z|X}}{1-\eta^2_{D\sim Z|X}}}{
\rho_P^2\times\eta^2_{P\sim Z|D,X}  \times \frac{\eta^2_{D\sim Z|X}}{1-\eta^2_{D\sim Z|X}}} \times
\frac{S_Y^2}{S_P^2} 
= k \times \frac{S_Y^2}{S_P^2}
\end{aligned}
\end{equation}

$k \overset{\Delta}{=} \frac{\rho_Y^2\times \eta^2_{Y\sim Z|D,X}  \times \frac{\eta^2_{D\sim Z|X}}{1-\eta^2_{D\sim Z|X}}}{
\rho_P^2\times \eta^2_{P\sim Z|D,X}  \times \frac{\eta^2_{D\sim Z|X}}{1-\eta^2_{D\sim Z|X}}}$. Which means that we can write Equation \ref{sss_e_part_lin_k}. We could use Equation \ref{sss_e_part_lin_k} for partial identification by reasoning about plausible ranges of values for $k$ and $\theta_{l,P}$, while estimating the remaining terms from observed data. $k$ represents the relative level of confounding where differences in scale have been accounted for. $\theta_{l,P}$ is the effect of $D$ on $P$. See \cite{Chernozhukov2022} for further discussion of $\eta^2$s, $S_Y^2$, $S_P^2$, $\rho_Y^2$, and $\rho_P^2$. 

\begin{equation} \label{sss_e_part_lin_k}
\begin{aligned} 
\theta_{l,Y}
&= 
\theta_{s,Y} - \text{sign}(m)\times \sqrt{k} \times (\theta_{s,P} - \theta_{l,P}) \times \sqrt{\frac{S_Y^2}{S_P^2}}
\end{aligned}
\end{equation}

\subsection{Non-parametric framework}

An investigator may also be interested in estimating a linear functional of the conditional expectation function of the outcome in a fully non-parametric setting, like $\theta_{l,Y} = \mathbb{E}[Y_1 - Y_0] = \mathbb{E}[f_Y(1,\mathbf{X},\mathbf{Z}) - f_Y(0,\mathbf{X},\mathbf{Z})]$ for a binary treatment $D$, where $Y_d = f_Y(d,\mathbf{X},\mathbf{Z},U_Y)$ is the equation for $Y$ in the structural causal model under intervention to set $D=d$. Again, the investigator is only able to estimate  $\theta_{s,Y} = \mathbb{E}[f_Y^*(1,\mathbf{X}) - f_Y^*(0,\mathbf{X})]$, where $f_Y^*(D,X)\overset{\Delta}{=}\mathbb{E}[Y|D,\mathbf{X}] = \mathbb{E}[f_Y(D,X,\mathbf{Z})|D,\mathbf{X}]$. We define $\theta_{l,P}$ and $\theta_{s,P}$ similarly for a placebo outcome. We could again consider an approach like Equation \ref{e_nonparam_m}. But again $m \overset{\Delta}{=} \frac{\text{bias}_{\text{(YD.X)}}}{\text{bias}_{\text{(PD.X)}}}$ may be difficult to interpret when $Y$ and $P$ have differing scales.

\begin{equation} \label{e_nonparam_m}
\begin{aligned} 
\theta_{l,Y}
&= 
\theta_{s,Y} - m\times (\theta_{s,P} - \theta_{l,P})
\end{aligned}
\end{equation}

\cite{Chernozhukov2022} also show that omitted variable bias in non-parametric setting can be written as an expression in terms of $\eta^2_{Y\sim Z|D,X}$ and a second term that, in the case of targeting $\theta_{l,Y} = \mathbb{E}[Y_1 - Y_0]$ with a binary treatment $D$, is the ``average gain in the conditional precision with which we predict $D$ by using $Z$ in addition to $X$,'' which is somewhat similar to $\eta^2_{D\sim Z|X}$. Specifically, we have the following. 

\begin{subequations} 
\begin{align}
|\theta_{s,Y} - \theta_{l,Y}|^2
= |{\text{bias}_{\text{(YD.X)}}}|^2 
&= \rho_Y^2 \times S_Y^2 \times \eta^2_{Y\sim Z|D,X}  \times \frac{1- R^2_{\alpha\sim\alpha_s}}{R^2_{\alpha\sim\alpha_s}} 
\\
|\theta_{s,P} - \theta_{l,P}|^2
= |{\text{bias}_{\text{(PD.X)}}}|^2 
&= \rho_P^2 \times S_P^2 \times \eta^2_{P\sim Z|D,X}  \times \frac{1- R^2_{\alpha\sim\alpha_s}}{R^2_{\alpha\sim\alpha_s}} 
\end{align}
\end{subequations}

$\text{BF}_{\text{(YD.X)}} \overset{\Delta}{=} \rho_Y^2 \times\eta^2_{Y\sim Z|D,X}\frac{1- R^2_{\alpha\sim\alpha_s}}{R^2_{\alpha\sim\alpha_s}} $ and  $\text{BF}_{\text{(PD.X)}} \overset{\Delta}{=} \rho_P^2 \times\eta^2_{P\sim Z|D,X}\frac{1- R^2_{\alpha\sim\alpha_s}}{R^2_{\alpha\sim\alpha_s}} $ are ``bias factors'' that are scaled versions of the unobserved confounding from $\mathbf{Z}$. 
$S_Y^2$ and $S_P^2$ are identified and therefore estimable from the data. See \cite{Chernozhukov2022} for discussion of $S_Y^2$, $S_P^2$, and $R^2_{\alpha\sim\alpha_s}$. 
Proceeding as in the last section, we can arrive at Equation \ref{sss_e_nonparam_k}. Here $k = \frac{\rho_Y^2 \times\eta^2_{Y\sim Z|D,X}\frac{1- R^2_{\alpha\sim\alpha_s}}{R^2_{\alpha\sim\alpha_s}} }{\rho_P^2 \times\eta^2_{P\sim Z|D,X}\frac{1- R^2_{\alpha\sim\alpha_s}}{R^2_{\alpha\sim\alpha_s}} }$ and again represents the relative level of confounding where differences in scale have been accounted for.

\begin{equation} \label{sss_e_nonparam_k}
\begin{aligned} 
\theta_{l,Y}
&= 
\theta_{s,Y} - \text{sign}(m)\times \sqrt{k} \times (\theta_{s,P} - \theta_{l,P}) \times \sqrt{\frac{S_Y^2}{S_P^2}}
\end{aligned}
\end{equation}

\section{Re-expression of bias factors}
\label{reexpress_BF}

We can rearrange the expression for omitted variable bias to obtain an expression for the ``bias factor'' (i.e., $R_{Y \sim Z | D,X}  f_{D\sim  Z|X}$). 

$$\begin{aligned}
\beta_{Y\sim D | X} - \beta_{Y\sim D| Z, X} 
&= R_{Y \sim Z | D,X}  f_{D\sim  Z|X} 
 \frac{\sd(Y^{\perp D,X})}{\sd(D^{\perp X})} \\
\implies 
R_{Y \sim Z | D,X}  f_{D\sim  Z|X}
&= (\beta_{Y\sim D | X} - \beta_{Y\sim D| Z, X} )
 \frac{\sd(D^{\perp X})}{\sd(Y^{\perp D,X})} \\
&= 
R_{Y\sim D | X} 
\frac{\sd(Y^{\perp X})}{\sd(D^{\perp X})}
\frac{\sd(D^{\perp X})}{\sd(Y^{\perp D,X})}
 - 
 R_{Y\sim D| Z, X}
\frac{\sd(Y^{\perp X,Z})}{\sd(D^{\perp X,Z})}
\frac{\sd(D^{\perp X})}{\sd(Y^{\perp D,X})}
\frac{\sd(Y^{\perp D,X,Z})}{\sd(Y^{\perp D,X,Z})} \\
&=
f_{Y\sim D | X} 
- 
f_{Y\sim D| Z, X}
\frac{\sd(Y^{\perp D,X,Z})}{\sd(D^{\perp X,Z})}
\frac{\sd(D^{\perp X})}{\sd(Y^{\perp D,X})} \\
&=
f_{Y\sim D | X} 
- 
f_{Y\sim D| Z, X}
\sqrt{\frac{1-R^2_{Y\sim Z|D,X}}{1-R^2_{D\sim Z|X}}} \\
\end{aligned}$$

This provides an expression for Cohen's f, $f_{Y\sim D | X}$ ($= \frac{R_{Y\sim D | X}}{\sqrt{1-R_{Y\sim D | X}^2}}$):\footnote{
We also note a nice symmetry: 
$1
=
\frac{f_{Y\sim D | X} }{R_{Y \sim Z | D,X}  f_{D\sim  Z|X}}
- 
\frac{f_{Y\sim D| Z, X} }{f_{Y \sim Z | D,X}  R_{D\sim  Z|X}}$.
}

$$\begin{aligned}
f_{Y\sim D | X} 
&= R_{Y \sim Z | D,X}  f_{D\sim  Z|X} + f_{Y\sim D| Z, X}
\sqrt{\frac{1-R^2_{Y\sim Z|D,X}}{1-R^2_{D\sim Z|X}}}
\end{aligned}$$

We can apply the omitted variable bias framework to $Y(d)$, the potential outcome for $Y$ where $D=d$, to understand the ``bias factor'' better. $Z,X$ are sufficient to de-confound the $D\to Y$ relationship, so $\beta_{Y(d)\sim D| Z, X} = f_{Y(d)\sim D| Z, X} =0$. Therefore, $f_{Y(d)\sim D | X} = R_{Y(d) \sim Z | D,X}  f_{D\sim  Z|X}$. 
Thus the ``bias factor'' for the potential outcome can be seen as a simple transformation (Cohen's f) of the correlation between $D$ and $Y(d)$, which comes only from the $D\leftarrow Z \to Y$ path. If we suppose that $Y = Y(d) = Y(d'), \forall d,d' \in \mathcal{D}$, where $\mathcal{D}$ is the support of $D$ (i.e., there is no causal effect of $D$ on $Y$), then $R_{Y(d) \sim Z | D,X} = R_{Y \sim Z | D,X}$ and the bias factor equals the Cohen's f for the potential outcome: $R_{Y \sim Z | D,X}  f_{D\sim  Z|X} = f_{Y(d)\sim D | X} $. In general, however, $R_{Y(d) \sim Z | D,X} \ne R_{Y \sim Z | D,X}$. A similar analysis can apply to placebos.

\newpage
\section{Example R Code}\label{code}

\subsection{Install PlaceboLM Package}

\begin{Verbatim}[frame=single]
# install.packages("devtools")
devtools::install_github("Adam-Rohde/PlaceboLM")
library(PlaceboLM)
\end{Verbatim}

\subsection{Placebo Treatment Simulation}

\begin{Verbatim}[frame=single]
# simulate data

set.seed(1)

n = 5000

ses  = runif(n,0,1)
phy  = rbinom(size = 1, n = n, prob = pnorm(2 * ses))
diet = rbinom(size = 1, n = n, prob = pnorm(-1 + ses))
HD   = rbinom(size = 1, n = n, prob = pnorm(-2 + -0.43*phy + 4*ses))

health_data = data.frame(
    SES = ses, 
    Physical_Activity = phy, 
    Dietary_Supplements = diet, 
    Heart_Disease = HD)
\end{Verbatim}


\begin{Verbatim}[frame=single]
# short regression (biased effect estimate)
summary(lm(Heart_Disease ~ Physical_Activity + Dietary_Supplements, 
data = health_data))

# long regression (true effect estimate)
summary(lm(Heart_Disease ~ Physical_Activity + Dietary_Supplements + SES, 
data = health_data))
\end{Verbatim}

\begin{Verbatim}[frame=single]
# placeboLM analysis 

placebolm.out = placeboLM(
  data = "health_data",
  outcome = "Heart_Disease",
  treatment = "Physical_Activity",
  placebo_treatment = "Dietary_Supplements",
  PY = "->",
  partialIDparam_minmax = list(k = c(-0.1,3), 
  coef_Y_P_given_DXZ = c(-0.5,0.5)) )

placeboLM_contour_plot(placebolm.out, gran = 100)

placeboLM_line_plot(placebolm.out, bootstrap=TRUE, n_boot=1000, 
ptiles = c(0.5), focus_param = "k", 
ptile_param = "coef_Y_P_given_DXZ", gran= 10, alpha = 0.01)

# We also show the true effect on the plots. 
\end{Verbatim}

\subsection{National Supported Work Demonstration}

\begin{Verbatim}[frame=single]
# get data
data(lalonde,package = "qte")
\end{Verbatim}

\begin{Verbatim}[frame=single]
# experimental estimate
summary(lm(re78 ~ treat , data = lalonde.exp))
# experimental estimate with covariates
summary(lm(re78 ~ treat + age + education + black + hispanic + married + 
nodegree, data = lalonde.exp))
\end{Verbatim}

\begin{Verbatim}[frame=single]
# placeboLM analysis 

placebolm.out = placeboLM(
  data = "lalonde.psid",
  outcome = "re78",
  treatment = "treat",
  placebo_outcome = "re74",
  observed_covariates = c("age", "education", "black", "hispanic", 
  "married", "nodegree"), partialIDparam_minmax = list(k = c(-0.1,2), 
  coef_P_D_given_XZ = c(-15000,15000)))

set.seed(0)
placeboLM_contour_plot(placebolm.out,gran= 100)

placeboLM_line_plot(placebolm.out,
                    bootstrap=TRUE,
                    n_boot=1000,
                    ptiles = c(0.5),
                    focus_param = "k",
                    ptile_param = "coef_P_D_given_XZ",
                    gran= 10,
                    alpha = 0.01)
                    
# We also show the experimental estimates on the plots. 
\end{Verbatim}


\begin{figure}[h!]
\begin{center}
\caption{\label{f_NSWD_LinePlot74} Line plot for NSW example using 1974 earnings as the placebo outcome. $\beta_{P\sim D| Z,X}=0$}
\includegraphics[scale=0.2]{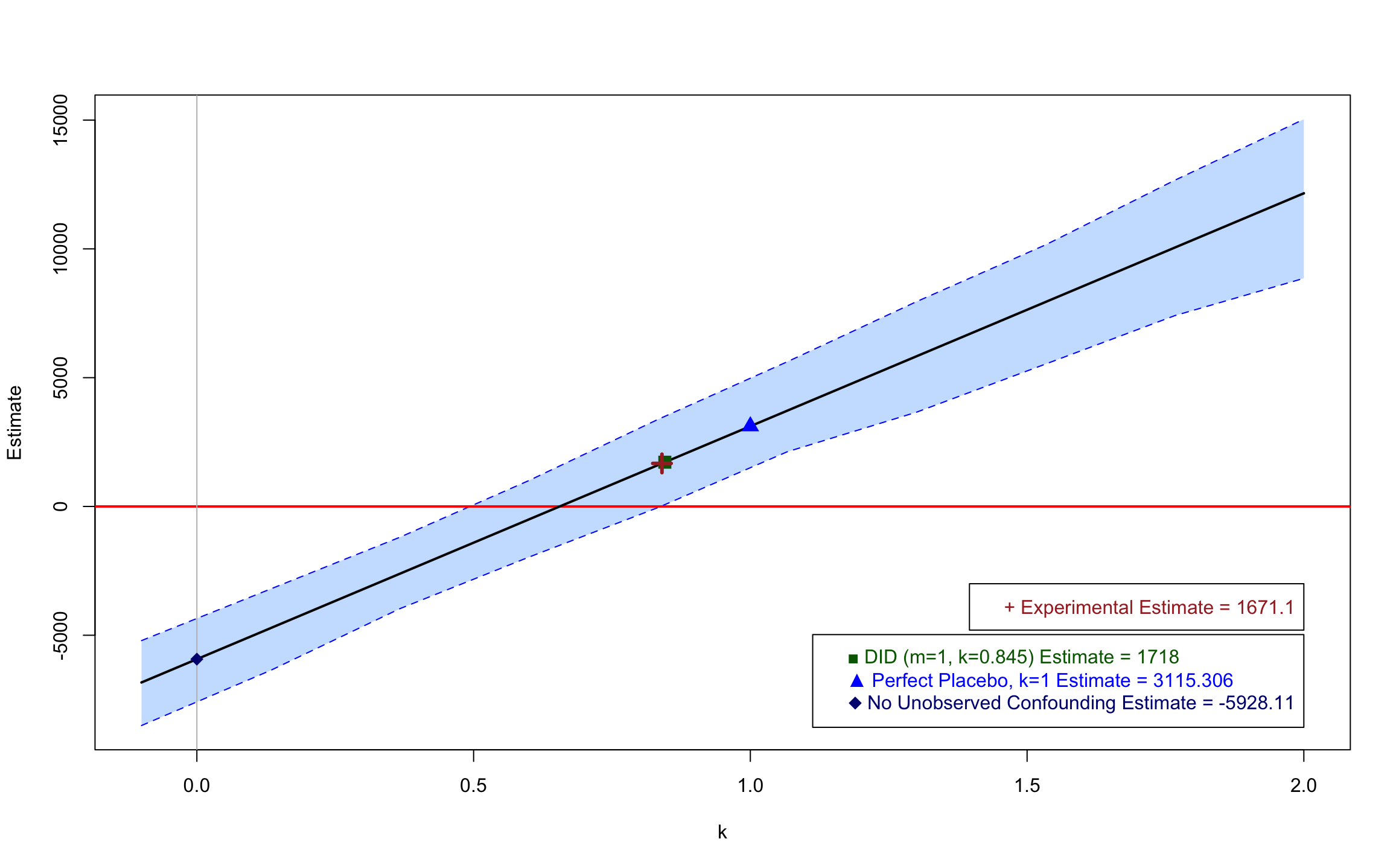}
\end{center}
\end{figure}

\subsection{Zika}

\begin{Verbatim}[frame=single]
# get data

library(haven)
library(tidyverse)

data = read_dta("zika_Table2.dta")

data_2014 = data %>%
  filter(StudyYear == 2014) %>%
  rename(Rate_2014 = Rate) %>%
  select(c(Code,Rate_2014,trt))

data_2016 = data %>%
  filter(StudyYear == 2016) %>%
  rename(Rate_2016 = Rate) %>%
  select(c(Code,Rate_2016,trt))

data_reshape = data_2014 %>%
  left_join(data_2016, by = join_by(Code,trt))
\end{Verbatim}

\begin{Verbatim}[frame=single]
# placeboLM analysis 

placebolm.out = placeboLM(
  data = "data_reshape",
  outcome = "Rate_2016",
  treatment = "trt",
  placebo_outcome = "Rate_2014",
  partialIDparam_minmax = list(k = c(0,2), coef_P_D_given_XZ = c(-10,10))
)


set.seed(0)

placeboLM_line_plot(placebolm.out,
                    bootstrap=TRUE,
                    n_boot=1000,
                    ptiles = c(0.5),
                    focus_param = "k",
                    ptile_param = "coef_P_D_given_XZ",
                    gran= 10,
                    alpha = 0.01)
\end{Verbatim}

\end{document}